\documentclass[aip,jcp,preprint]{revtex4-2}
\usepackage[latin2]{inputenc}
\usepackage[english]{babel}
\usepackage{t1enc}
\usepackage{amsfonts}
\usepackage{graphicx}
\usepackage{amssymb}
\usepackage{amsmath}
\usepackage{textcomp}
\usepackage{morefloats}
\usepackage{gensymb}
\usepackage{fullpage}
\usepackage{enumerate}
\usepackage{rotating}
\usepackage{acronym}
\usepackage{scalefnt}
\usepackage{epsfig}
\usepackage{enumitem}
\usepackage{wrapfig}
\usepackage{esint}
\usepackage{float}
\usepackage{upgreek}
\usepackage{bm}
\usepackage{amsbsy}
\usepackage{color}

\makeatletter
\providecommand\barcirc{\mathpalette\@barred\circ}
\def\@barred#1#2{\ooalign{\hfil$#1-$\hfil\cr\hfil$#1#2$\hfil\cr}}

\makeatother

\renewcommand\footnotemark{}

\newcommand{\mat}[1]{{\bf #1}}

\begin{document}

\title{Impact of Cavity on Molecular Ionization Spectra}

\author{Csaba F\'abri}
\email{ficsaba@staff.elte.hu}
\affiliation{HUN-REN--ELTE Complex Chemical Systems Research Group, H-1518 Budapest 112, Hungary}
\affiliation{Department of Theoretical Physics, University of Debrecen, P.O. Box 400, H-4002 Debrecen, Hungary}

\author{G\'abor J. Hal\'asz}
\affiliation{Department of Information Technology, University of Debrecen, P.O. Box 400, H-4002 Debrecen, Hungary}

\author{Lorenz S. Cederbaum}
\affiliation{Theoretische Chemie, Physikalisch-Chemisches Institut, Universit\"at Heidelberg, D-69120, Germany}

\author{\'Agnes Vib\'ok}
\email{vibok@phys.unideb.hu}
\affiliation{Department of Theoretical Physics, University of Debrecen, P.O. Box 400, H-4002 Debrecen, Hungary}
\affiliation{ELI-ALPS, ELI-HU Non-Profit Ltd, H-6720 Szeged, Dugonics t\'er 13, Hungary}

\begin{abstract}
Ionization phenomena are widely studied for decades. With the advent of cavity technology, the question arises how quantum light affects molecular ionization. 
As the ionization spectrum is recorded from the neutral ground state, it is usually possible to choose cavities which exert negligible effect on the neutral ground state, 
but have significant impact on the ion and the ionization spectrum. Particularly interesting are cases where the ion exhibits conical intersections between close-lying 
electronic states, which gives rise to substantial nonadiabatic effects. 
Assuming single-molecule strong coupling,
we demonstrate that vibrational modes irrelevant in the absence of cavity play a decisive role when 
the molecule is in the cavity. Here, dynamical symmetry breaking is responsible for the ion-cavity coupling and high symmetry enables control of the coupling via molecular 
orientation relative to the cavity field polarization. Significant impact on the spectrum by the cavity is found and shown to even substantially increase for less symmetric molecules.
\end{abstract}

\maketitle

Molecular cavity quantum electrodynamics studies the interaction of confined electromagnetic field modes with molecules.
Photon-molecule coupling gives rise to mixed light-matter states which are called polaritons carrying both
photonic and molecular features. Since the pioneering experimental work of the Ebbesen group reported in 2012,\cite{Hutchison2012}
``polaritonic chemistry" has become a rapidly emerging field of physics and chemistry opening up novel possibilities to manipulate material properties.
An array of intriguing experimental and theoretical works have demonstrated that polaritonic states can dramatically alter physical and chemical properties of molecular systems. \cite{Ebbesen20162403,Chikkaraddy2016127,Kowalewski20162050,Flick20173026,Feist2018205,Fregoni2018,Ribeiro20186325,Szidarovszky20186215,Ruggenthaler2018,Vendrell2018,Csehi2019,Csehi2019a,Reitz2019,Ulusoy20198832,Mandal2019,Triana2019,19YuMe,Davidsson2020234304,Fabri2020234302,Gu20201290,Gu2020a,Silva2020,Felicetti20208810,Fregoni2020,Mandal2020,20PoJaLy,Fabri2021a,Garcia-Vidal2021,Huang2021,Szidarovszky2021,Cederbaum2021,Cederbaum2021a,Schfer2021,Triana2021,21FiSa,D1CP00943E,Fbri2022,Schfer2022,Fischer2022,Li2022,Cui2022,Wellnitz2022,Malave2022,22ReSoGe,Fregoni2022,22BiHa,Fan2023,23ScKo,23KoDuPa,20ScRuRo,20MaMoHu,20TaMaZh}
Strong coupling has been shown to influence chemical reactivity by enhancing or suppressing available mechanisms
\cite{Hutchison2012,Galego2016}, and mediating new ones.\cite{Mandal2019}
Strong coupling can also enhance charge and energy transfer,\cite{Mandal2020} modify absorption spectra \cite{Szidarovszky20186215,Szidarovszky2020,Fabri2020234302,Fabri2021a}
and give rise to strong nonadiabatic effects in molecules \cite{Szidarovszky20186215,Feist2018205,Fregoni2018,Csehi2019,Fregoni2020,Gu2020a,Gu20201290,Fabri2020234302,Fabri2021a,D1CP00943E,Szidarovszky2021,Fbri2022,Fischer2022,Fbri2023}
by providing ultrafast nonradiative decay channels.\cite{Vendrell2018,Ulusoy20198832,Csehi2019a,Csehi2022}

Coupling between nuclear and electronic motions in polyatomic molecules induces nonadiabatic phenomena, such as
conical intersections (CIs).\cite{Koppel198459,Yarkony1996985,Baer_2002,Domcke2004,Worth2004127,Baer2006}
CIs between electronic potential energy surfaces (PESs) result in remarkable changes in the dynamical, spectroscopic and topological
properties of molecules. Additionally, nonadiabatic effects can also be created by external classical or quantized electromagnetic
fields.\cite{Sindelka2011,Halasz2015348,Kowalewski20162050,Szidarovszky20186215,Feist2018205,Fabri2021a}
In such cases, the laser field or a confined mode of the cavity can couple molecular electronic states and light-induced CIs
(LICIs) are formed. Nonadiabatic effects associated with LICIs are essentially identical to natural ones. 

We investigate the combined impact of natural and light-induced CIs on the ionization spectrum of a molecule in a cavity.
Although natural and light-induced nonadiabatic phenomena \cite{Badanko2022,Csehi2019,Csehi2019a,Csehi2022,D1CP00943E,Feist2018205,Fregoni2018,Fregoni2020,Galego2016,Gu20201290,Gu2020a,Triana2019,Ulusoy20198832,Fbri2022,Fbri2023}
and their signatures in absorption spectra \cite{Szidarovszky20186215,Szidarovszky2020,Szidarovszky2021,Fabri2020234302,Fabri2021a}
have been examined in neutral molecules placed into a cavity, the ionization of molecules inside a cavity has remained largely unexplored
(Refs. \onlinecite{21DePrince} and \onlinecite{22RiHaRo} investigated the ionization potential of molecules in cavity).
To fill this gap, we choose the butatriene (BT, C$_4$H$_4$) molecule as a showcase system.
Since the low-energy (cavity-free) ionization spectrum of BT already exhibits a dramatic fingerprint of a natural
CI,\cite{Cederbaum1977,Cattarius2001} BT is a compelling candidate for the current study.
In particular, a natural CI is formed between the electronic ground ($\textrm{X} ~ ^2B_{2\textrm{g}}$) and first
excited states ($\textrm{A} ~ ^2B_{2\textrm{u}}$) of the BT cation (BT$^+$).
The CI is located near the Franck--Condon (FC) region of the neutral BT ground state and gives rise to an unexpected and
well-separated band (``mystery band") in the ionization spectrum.\cite{Brogli1974}
The origin of the mystery band was clarified in Ref. \onlinecite{Cederbaum1977} where it was also concluded that a vibronic coupling model
treating two vibrational modes is capable of accurately reproducing the low-energy experimental ionization spectrum.
Later, an all-mode vibronic coupling model was developed and gave results essentially identical to the 2-mode model.\cite{Cattarius2001}

Coupling to cavity leads to LICI formation and the ionization spectrum is shaped by the joint effect of natural and light-induced CIs.
In sharp contrast to natural CIs, the position of the LICI and the light-induced nonadiabatic coupling strength can
be controlled by the cavity frequency and coupling strength, respectively. This allows for a systematic
control of light-induced nonadiabaticity including the competition between natural and light-induced CIs.
The scenario of the current work is outlined as follows. BT is ionized with a weak laser pulse in a low-frequency cavity.
In neutral BT the ground and first excited electronic states are well separated energetically ($\sim 5.7 ~ \textrm{eV}$ at the FC point). 
Therefore, a low-frequency cavity mode tailored to bring the ground and first excited states of BT$^+$ into resonance does not couple the neutral BT ground state to other states.
However, resonant coupling of the X and A states of BT$^+$ leads to the formation of LICIs.
Consequently, significant cavity-induced changes in the ionization spectrum, also affecting the mystery band, are expected compared to the cavity-free case.
We shall see that BT is particularly interesting for studying nonadiabatic effects. Due to symmetry, the cavity does not couple to BT$^+$ at the FC point.
All couplings are induced dynamically by symmetry-breaking vibrational modes. Symmetry also allows to control which kind of modes couple. 

A single molecule coupled to a lossless cavity mode is described by the Hamiltonian \cite{04CoDuGr}
\begin{equation}
	\hat{H}_\textrm{cm} = \hat{H}_0 + \hbar \omega_\textrm{cav} \hat{a}^\dag \hat{a} - g \hat{\vec{\mu}} \vec{e} (\hat{a}^\dag + \hat{a})
   \label{eq:Hcm}
\end{equation}
where $\hat{H}_0$ is the molecular Hamiltonian, $\omega_\textrm{cav}$ denotes the cavity angular frequency,
$\hat{a}^\dag$ and $\hat{a}$ are creation and annihilation operators, $g$ refers to the coupling strength parameter,
$\hat{\vec{\mu}}$ corresponds to the molecular electric dipole moment and $\vec{e}$ is the field polarization vector.
In this work, we treat a molecule coupled to a plasmonic cavity mode
and omit the dipole self-energy (DSE) term 
(see Refs. \onlinecite{Fregoni2022}, \onlinecite{19GaClGa} and \onlinecite{21FeFeGa} 
for further explanation).
It is worth noting that strong coupling has been achieved experimentally at the
single-molecule (emitter) level in plasmonic cavities.\cite{Chikkaraddy2016127,22BiHa}
The relevance of the DSE, including its role in collective vibrational strong
coupling, has been investigated thoroughly.\cite{18RoWeRu,20ScRuRo}
It was also concluded that without the DSE
the coupled cavity-molecule system does not have a 
ground state.\cite{18RoWeRu} We stress that in the
current computational model a stable ground state was
found without the DSE term. Inclusion of the DSE would
also modify our symmetry arguments, for example, the
DSE term would give a nonzero contribution at the FC point.
Moreover, it is important to note that
plasmonic cavity modes have a short lifetime.\cite{Fregoni2022}
However, in the current case the finite lifetime of the
cavity mode primarily influences the shape of the spectral
lines (i.e., the spectral lines would have
a certain width) in the ionization spectrum.
Since we are interested in identifying cavity-induced
effects on line positions and intensities,
cavity loss has been omitted from our computational
protocol in this first investigation of the subject.

Considering two electronic states (X and A) of BT$^+$, the Hamiltonian of Eq. \eqref{eq:Hcm} reads
\begin{equation}
    \hat{H}_\textrm{cm}  = 
         \begin{bmatrix}
            \hat{T} + V_\textrm{X} & V_\textrm{XA} & W_{\textrm{X}}^{(1)} & W_{\textrm{XA}}^{(1)} & \dots \\
            V_\textrm{XA} & \hat{T} + V_\textrm{A} & W_{\textrm{XA}}^{(1)} & W_{\textrm{A}}^{(1)} & \dots \\
            W_{\textrm{X}}^{(1)} & W_{\textrm{XA}}^{(1)} & \hat{T} + V_\textrm{X} + \hbar\omega_\textrm{cav} & V_\textrm{XA} & \dots \\
            W_{\textrm{XA}}^{(1)} & W_{\textrm{A}}^{(1)} & V_\textrm{XA} &\hat{T} + V_\textrm{A} + \hbar\omega_\textrm{cav} & \dots \\
            \vdots & \vdots & \vdots & \vdots & \ddots 
        \end{bmatrix}
    \label{eq:cavity_H}
\end{equation}
where $\hat{T}$ is the kinetic energy operator, $V_\textrm{X}$ and $V_\textrm{A}$ are the ground-state and excited-state PESs, and
$V_\textrm{XA}$ describes the vibronic coupling between X and A in the diabatic representation.
The cavity-molecule coupling is characterized by $W_\alpha^{(n)} = -g \sqrt{n} \mu_\alpha$ with $\alpha = \textrm{X}, \textrm{A}$
and $W_{\textrm{XA}}^{(n)} = -g \sqrt{n} \mu_\textrm{XA}$ ($n = 0,1,2,\dots$ labels Fock states of the cavity mode).
In practical computations we have included all counter-rotating terms
and employed the maximal photon numbers $n_\textrm{max} = 15$
(for polarizations $\mathbf{e}=(1,0,0)$ and $\mathbf{e}=(0,1,0)$) and $n_\textrm{max} = 20$
(for $\mathbf{e}=(1,1,0)/\sqrt{2}$).
The permanent (PDM) and transition (TDM) dipole moment components along $\vec{e}$ are denoted by
$\mu_\alpha$ ($\alpha = \textrm{X}, \textrm{A}$) and $\mu_\textrm{XA}$, respectively.
We stress that the rotational degrees of freedom are omitted in the current work
and the orientation of the molecule is kept fixed with respect to $\vec{e}$.
Polaritonic (adiabatic) PESs can be obtained by diagonalizing the potential energy part of $\hat{H}_\textrm{cm}$.

Vibronic coupling models have been extremely successful in describing ionization spectra.\cite{Koppel198459,17WeChKi}
Accordingly, the potentials  $V_\textrm{X}$, $V_\textrm{A}$ and $V_\textrm{XA}$, and also the terms $W_{\textrm{X}}^{(n)}$, $W_{\textrm{A}}^{(n)}$
and $W_{\textrm{XA}}^{(n)}$,  are expanded around the FC point of $D_{2\textrm{h}}$ point-group symmetry.
In the 2-mode vibronic coupling model of BT$^+$,\cite{Koppel198459,Cederbaum1977,Cattarius2001}
the two electronic states are coupled by the torsional mode of $A_\textrm{u}$ symmetry (coupling mode) and
the energy gap between them is tuned by the central C-C stretch mode of $A_\textrm{g}$ symmetry (tuning mode). 
Vibrational modes relevant for the current study are listed in Table II of the Supporting Information and see Fig. 1 there for body-fixed axis definitions.
The coupling ($\nu_\textrm{c}$) and tuning ($\nu_\textrm{t}$) modes correspond to $\nu_5$ and $\nu_{14}$, respectively. Following earlier work,\cite{Cederbaum1977,Cattarius2001}
the potentials are approximated as 
\begin{gather}
	V_\alpha = \epsilon_\alpha + 
    \frac{1}{2} \omega_\textrm{t}^2 Q_\textrm{t}^2 + 
    \frac{1}{2} \omega_\textrm{c}^2 Q_\textrm{c}^2 +
    \kappa_\alpha Q_\textrm{t} \nonumber \\
    V_\textrm{XA} = \lambda Q_\textrm{c}
\end{gather}
where $\alpha=\textrm{X}, \textrm{A}$,
$Q_\textrm{t}$ and $Q_\textrm{c}$ denote
normal coordinates of the tuning and coupling modes,
respectively, while $\omega_\textrm{t}$ and 
$\omega_\textrm{c}$ refer to ground-state
vibrational frequencies of BT. For a more detailed
description of the 2-mode model and parameter values
we refer to Sections II and IV/A of the Supporting Information.

Group-theoretical considerations (see Section I of the Supporting Information)
reveal that the TDM and PDMs all vanish at the FC point.
Moreover, the TDM and PDMs remain zero upon moving away from the FC point along $\nu_\textrm{c}$ and $\nu_\textrm{t}$.
In other words, the cavity does not couple to BT$^+$ in the 2-mode model and in order to enable this coupling we incorporate two additional modes,
one which produces TDM ($\nu_9$) and one which produces PDMs ($\nu_{10}$) upon displacement from the FC point, breaking the $D_{2\textrm{h} }$ symmetry.
We shall denote these modes by $\nu_{_\textrm{TDM}}$ and $\nu_{_\textrm{PDM}}$ and address the resulting model consisting of these modes necessary
to couple the cavity and the molecule and the tuning and coupling modes necessary to describe the natural CI, as the 4-mode model.
Obviously, the choice of $\nu_{_\textrm{TDM}}$ and $\nu_{_\textrm{PDM}}$ is not unique. In the current work, modes $\nu_{_\textrm{TDM}}$ and
$\nu_{_\textrm{PDM}}$ were selected to generate appreciable TDM and PDM values along two orthogonal axes (see the discussion later).
The PDM and TDM functions are expressed
as $\mu_\alpha = \beta_\alpha Q_{_\textrm{PDM}}$ and
$\mu_\textrm{XA} = \gamma Q_{_\textrm{TDM}}$, respectively,
where $\alpha=\textrm{X}, \textrm{A}$, and
$Q_{_\textrm{PDM}}$ and $Q_{_\textrm{TDM}}$ are
normal coordinates of the PDM and TDM modes.
Values of the parameters $\beta_\alpha$ and $\gamma$ as
well as further information on the construction of the
PDM and TDM functions can be found in Sections IV/A and
IV/B of the Supporting Information.
For the evaluation of Hamiltonian matrix elements
and computation of ionization spectra we refer to
Sections II and III of the Supporting Information.

\begin{figure*}
\includegraphics[scale=0.75]{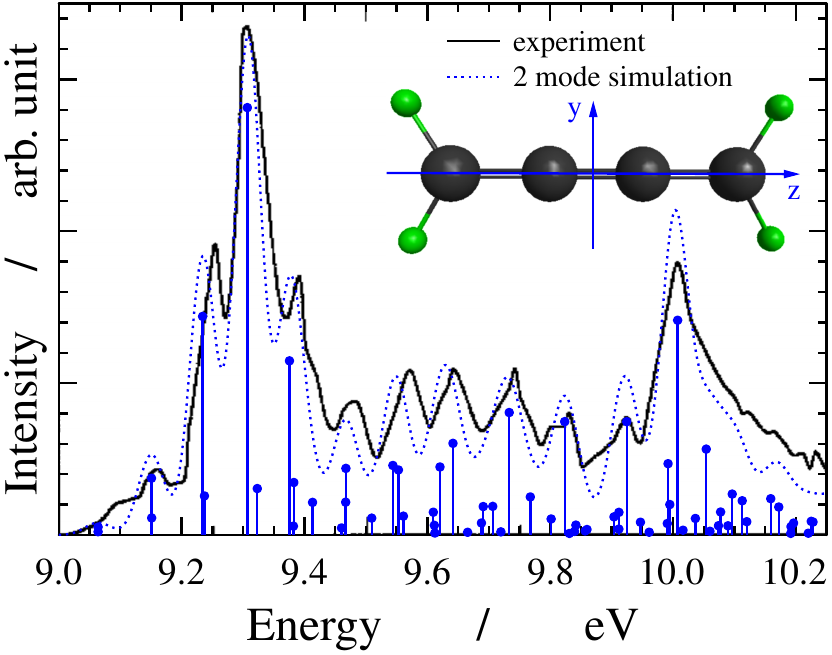}
\caption{\label{fig:1} Experimental and calculated (2-mode model) ionization spectra of butatriene (BT), 
showing nice agreement.
See the inset for the equilibrium structure of neutral BT (and body-fixed coordinate axis definitions),
corresponding to the Franck--Condon point. 
The calculated spectral lines are convoluted with a Gaussian function of full width at half maximum of 
$0.05 ~ \textrm{eV}$ to account for the experimental resolution.}
\end{figure*}

Fig. \ref{fig:1} shows the experimental \cite{Brogli1974} and calculated (2-mode) ionization spectra of BT.
The two spectra show a nice agreement which validates the 2-mode model.\cite{Cederbaum1977,Cattarius2001}
The equilibrium structure of BT (FC point) and body-fixed axis definitions are also visible
in Fig. \ref{fig:1}.
We mention that the interaction between the cavity field and 
the emitted electron is neglected in the current work.
This approximation is justified by the fact that the
emitted electron is rather fast (its kinetic energy exceeds
$10 ~ \textrm{eV}$ in the experimental spectrum shown 
in Fig. \ref{fig:1}). On the other hand, investigating 
the impact of the cavity-electron interaction on the 
ionization spectrum is of interest by itself, which is left for future work.

\begin{figure*}
\includegraphics[width=0.8\textwidth]{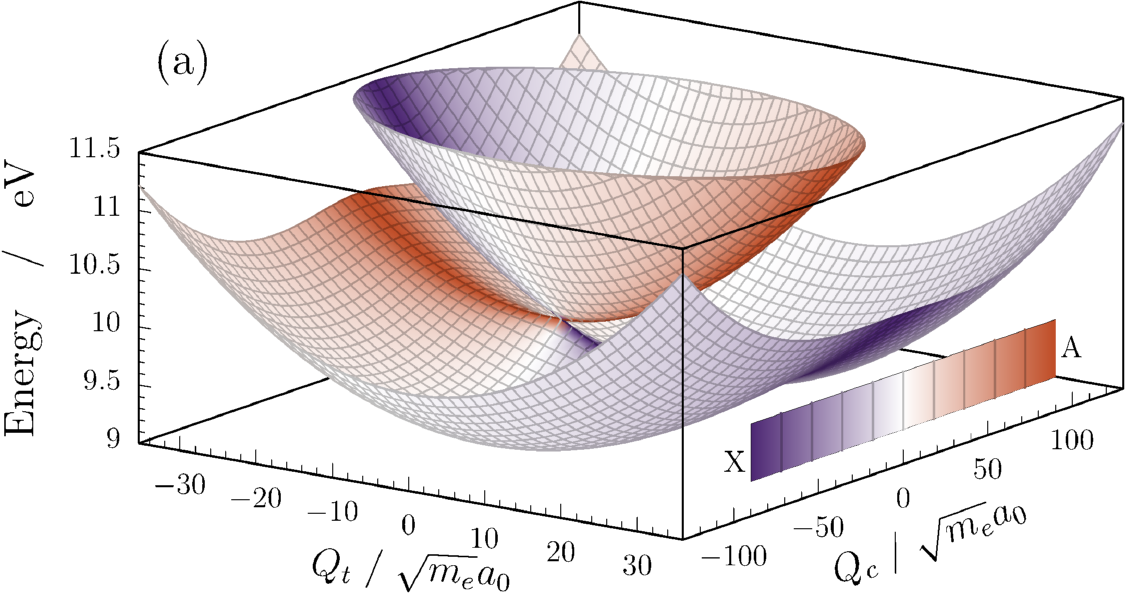}\\
\includegraphics[width=0.8\textwidth]{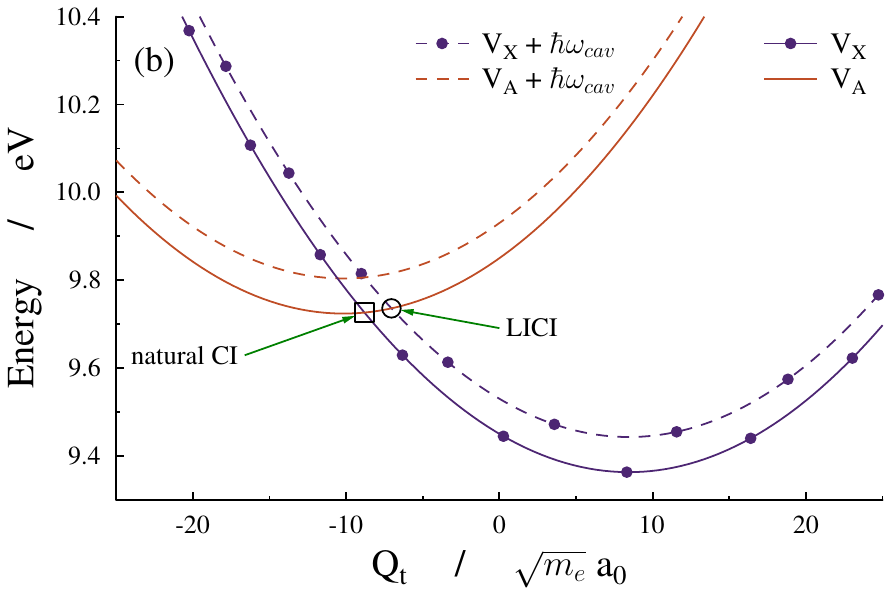}
\caption{\label{fig:2}
		(a) Two-dimensional adiabatic (cavity-free) potential energy surfaces (PESs) of the butatriene cation (BT$^+$) along the
		      tuning ($Q_\textrm{t}$) and coupling ($Q_\textrm{c}$) modes.
		(b) One-dimensional PES cuts of BT$^+$ along $Q_\textrm{t}$ with $Q_\textrm{c} = Q_{_\textrm{TDM}} = Q_{_\textrm{PDM}} = 0$.
		      $V_\textrm{X}$ and $V_\textrm{A}$ refer to the ground-state and excited-state PESs of BT$^+$, while
		      $V_\textrm{X} + \hbar \omega_\textrm{cav}$ and $V_\textrm{A} + \hbar \omega_\textrm{cav}$ are the same curves
		      shifted by the photon energy with $\omega_\textrm{cav} = 645.02 ~ \textrm{cm}^{-1}$
            ($\sim 0.08 ~ \textrm{eV}$).
		      Positions of the natural conical intersection (CI, between $V_\textrm{X}$ and $V_\textrm{A}$) and the light-induced conical intersection
		      (LICI, between $V_\textrm{X} + \hbar \omega_\textrm{cav}$ and $V_\textrm{A}$) are explicitly marked in the figure.}
\end{figure*}

Fig. \ref{fig:2} shows different PES cuts. It is clearly visible in panel a of Fig. \ref{fig:2} that the two-dimensional adiabatic (cavity-free)
PESs of BT$^+$ along normal coordinates $Q_\textrm{t}$ and $Q_\textrm{c}$ form a natural CI at $78447 ~ \textrm{cm}^{-1}$ ($9.73 ~ \textrm{eV}$) 
above the minimum of the BT ground state. Panel b of Fig. \ref{fig:2} provides one-dimensional BT$^+$ PES cuts along $Q_\textrm{t}$
($Q_\textrm{c} = Q_{_\textrm{TDM}} = Q_{_\textrm{PDM}} = 0$). Besides $V_\textrm{X}$ and $V_\textrm{A}$ one can also see $V_\textrm{X} + \hbar \omega_\textrm{cav}$ and
$V_\textrm{A} + \hbar \omega_\textrm{cav}$ in panel b, both shifted by the photon energy with $\omega_\textrm{cav} = 645.02 ~ \textrm{cm}^{-1}$ ($\sim 0.08 ~ \textrm{eV}$).
As indicated in panel b, the natural CI at $Q_\textrm{t} = -8.80 ~ \sqrt{m_\textrm{e}} a_0$ (between $V_\textrm{X}$ and $V_\textrm{A}$) is adjacent to
the LICI at  $Q_\textrm{t} = -7.04 ~ \sqrt{m_\textrm{e}} a_0$ (between $V_\textrm{X} + \hbar \omega_\textrm{cav}$ and $V_\textrm{A}$).

\begin{figure*}
\includegraphics[scale=0.575]{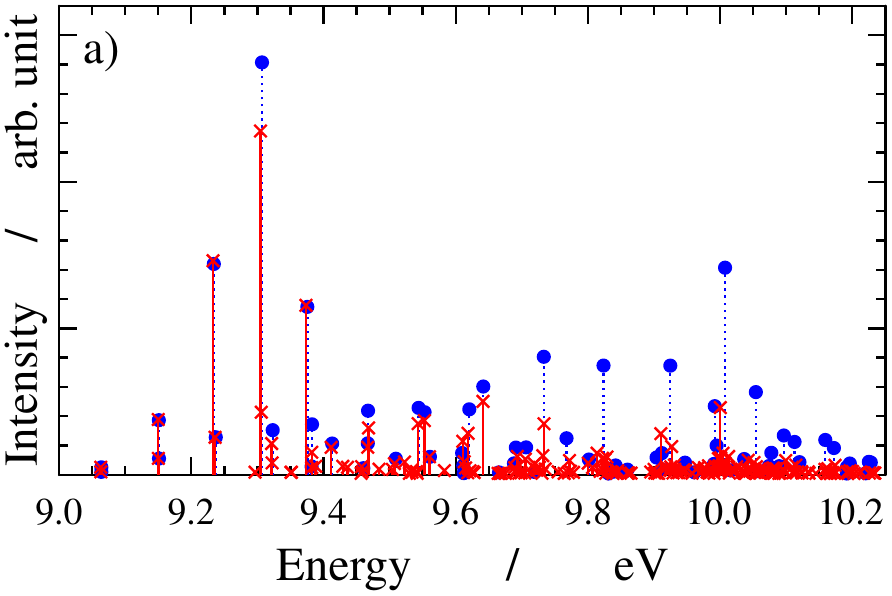}
\includegraphics[scale=0.575]{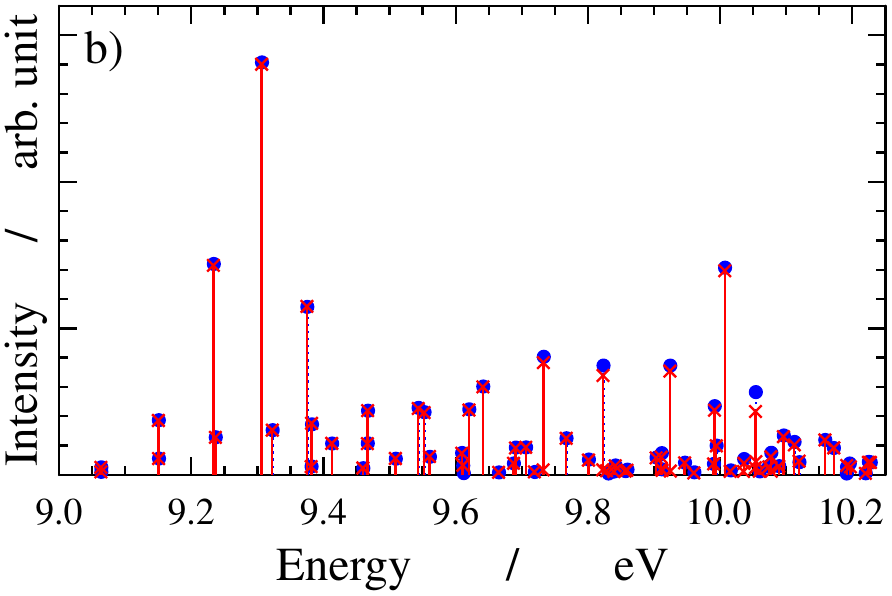}
\includegraphics[scale=0.575]{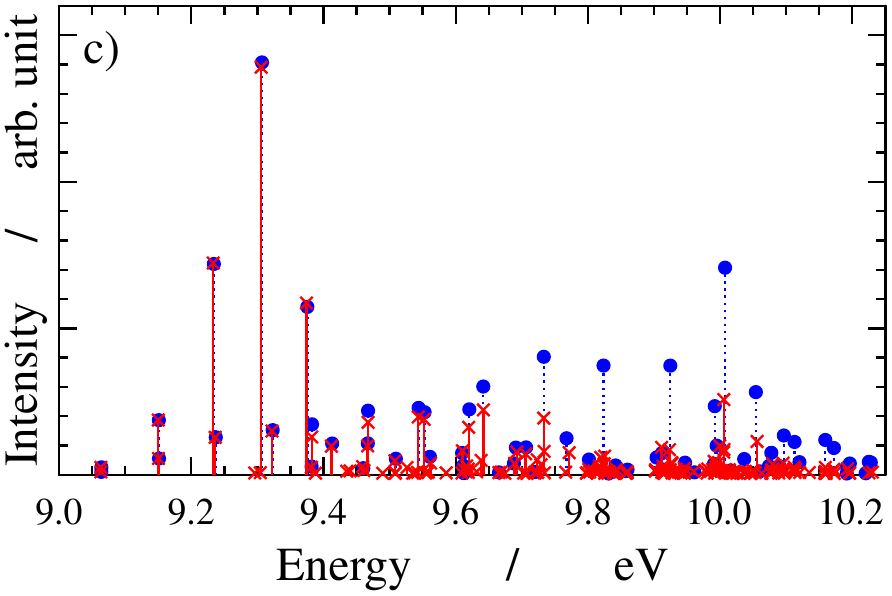}
\caption{\label{fig:3}
        Cavity-free (blue, $\bullet$) and cavity (red, x) ionization spectra of butatriene with cavity parameters 
        $\omega_\textrm{cav} = 645.02 ~ \textrm{cm}^{-1}$ and $g = 0.1 ~ \textrm{au}$.
        The field polarization vector is chosen in three different ways:
		 $\mathbf{e} = (0,1,0)$ (panel a, only permanent dipole moments (PDMs)), 
		 $\mathbf{e} = (1,0,0)$ (panel b, only transition dipole moment (TDM)), and
		 $\mathbf{e} = (1,1,0)/\sqrt{2}$ (panel c, both TDM and PDMs).
	Significant cavity-induced dynamical effects can be observed mainly in the nonadiabatic region
	(above $9.55 ~ \textrm{eV}$) of the ionization spectrum.}
\end{figure*}

Fig. \ref{fig:3} shows computed ionization spectra for $\omega_\textrm{cav} = 645.02 ~ \textrm{cm}^{-1}$ and $g = 0.1 ~ \textrm{au}$ (used for visualization purposes only) for the 4-mode model.
Results for three different polarizations of the cavity field are depicted in red in Fig. \ref{fig:3}. 
For comparison, ionization spectra of the 4-mode model, but without coupling to the cavity are also
shown (blue). The polarization in panel a is $\mathbf{e} = (0,1,0)$, i.e., along the $y$ axis. In this case, only the PDMs in the electronic states couple the molecule 
to the cavity via the $\nu_{_\textrm{PDM}}$ mode. In panel b the polarization is along the $x$ axis, $\mathbf{e} = (1,0,0)$, and the coupling is by the $\nu_{_\textrm{TDM}}$ mode only.
The orthogonality of the TDM to the PDMs enables to control the cavity-induced effects by changing the field polarization and to study the impact of the individual modes on the spectrum.
This has been another reason to choose BT for our investigation. The joint impact of the PDM and TDM modes can be studied by varying the polarization in the $xy$-plane,
an example is shown in panel c for $\mathbf{e} = (1,1,0)/\sqrt{2}$.
In each case, the system is initially assumed to be in the vibrational ground state of BT with zero photons in the cavity
which corresponds to the lowest eigenstate of the coupled cavity-molecule system to a very good approximation.

As already stated, the 2-mode vibrational model is insufficient to account for cavity-molecule interactions due to symmetry.
However, by making displacements along modes $\nu_{_\textrm{TDM}}$ and $\nu_{_\textrm{PDM}}$
the TDM and PDM values are no longer zero and consequently, BT$^+$ can interact with the cavity mode.
More precisely, the coupling to the cavity is induced by dynamical symmetry breaking activated by ionization in the cavity. 
The impact of the cavity on the ionization process strongly depends on the orientation of the molecule with respect to the field polarization.
Accordingly, besides the original $\nu_\textrm{c}$ coupling and $\nu_\textrm{t}$ tuning modes, $\nu_{_\textrm{PDM}}$
(only PDM, panel a of Fig. \ref{fig:3}), $\nu_{_\textrm{TDM}}$ (only TDM, panel b), or both $\nu_{_\textrm{TDM}}$ and $\nu_{_\textrm{PDM}}$ (panel c) come into play.
It is conspicuous in Fig. \ref{fig:3} that the cavity-free (blue) and cavity (red) spectra can differ considerably from each other as seen in panels a (only PDM) and c.
The impact of $\nu_{_\textrm{TDM}}$ (panel b) is found to be rather minor in BT$^+$. These observations are attributed to the strong mixing by the PDMs of the 
vibrational levels of the surfaces originating from the solid and dashed blue curves and those originating from the respective red curves in Fig. \ref{fig:2}.
The resulting hybrid light-matter states are subject to the electronic mixing imposed by the natural CI and are, therefore, expected to strongly change the
nonadiabatic effects found in the cavity-free case. Indeed, it is seen in panel a that intense lines in the cavity-free spectrum above about $9.55 ~ \textrm{eV}$
(blue) are split into many lines of tiny intensities (red). 

The energetic position of the natural CI is $9.55 ~ \textrm{eV}$ above the zero-point energy of BT.
Thus, following Ref. \onlinecite{Koppel198459}, the original cavity-free spectrum can be divided into adiabatic ($E<9.55 ~ \textrm{eV}$) and nonadiabatic ($E>9.55 ~ \textrm{eV}$) regions.
It is apparent in Fig. \ref{fig:3} that the nonadiabatic region is significantly modified by the cavity,
while the adiabatic region remains largely unaffected by cavity-molecule interactions (some levels mix in, but their energy splittings are minute).
This finding can be rationalized as follows.
Nonzero cavity-molecule couplings and cavity effects in the ionization spectrum can be ascribed to dynamical symmetry breaking
induced by displacement along modes $\nu_{_\textrm{TDM}}$ and $\nu_{_\textrm{PDM}}$.
Owing to the conditions required by the LICI formation (degenerate diabatic potentials and zero cavity-molecule couplings)
the LICI is situated near the natural CI for $\omega_\textrm{cav} = 645.02 ~ \textrm{cm}^{-1}$ (see panel b of Fig. \ref{fig:2}).
As a result, the natural CI and the LICI start exerting their effects roughly above the same energy and imprint their signatures in the nonadiabatic
region of the cavity-free ionization spectrum, including the mystery band which emerges due to natural nonadiabatic effects.
Several additional ionization spectra have been computed, supporting our conclusions.
Two of them are shown as examples in Section V of the Supporting Information.

We stress that the natural CI appears again between the surfaces originating from the two dashed curves and the same holds for the LICI
which appears again between the red dashed and the blue solid curves (see Fig. \ref{fig:2}). This second ``natural CI" is actually also a light-induced CI as it
does not exist without the cavity. Owing to the low cavity frequency employed, nonadiabatic effects discussed above are due to all of these four conical intersections. 

\begin{figure*}
\includegraphics[scale=0.65]{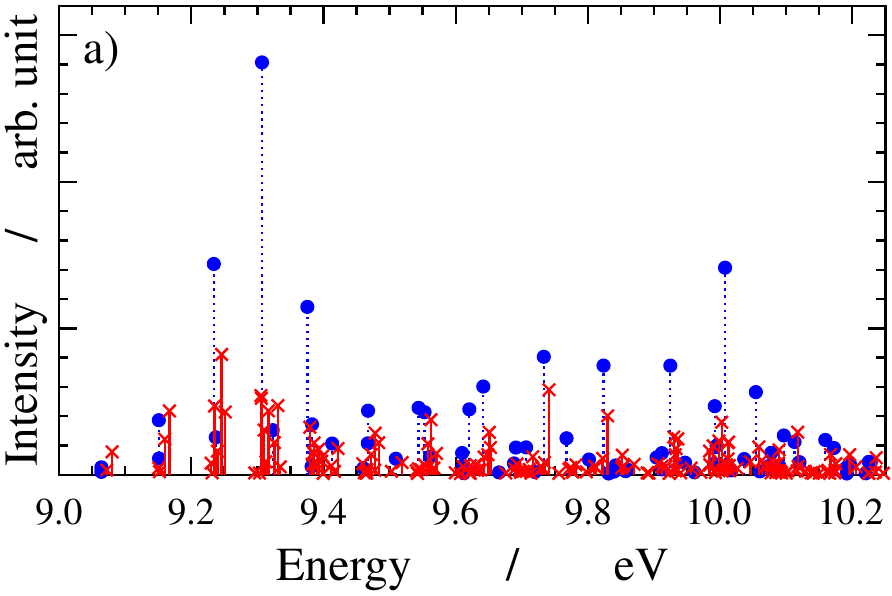}
\includegraphics[scale=0.65]{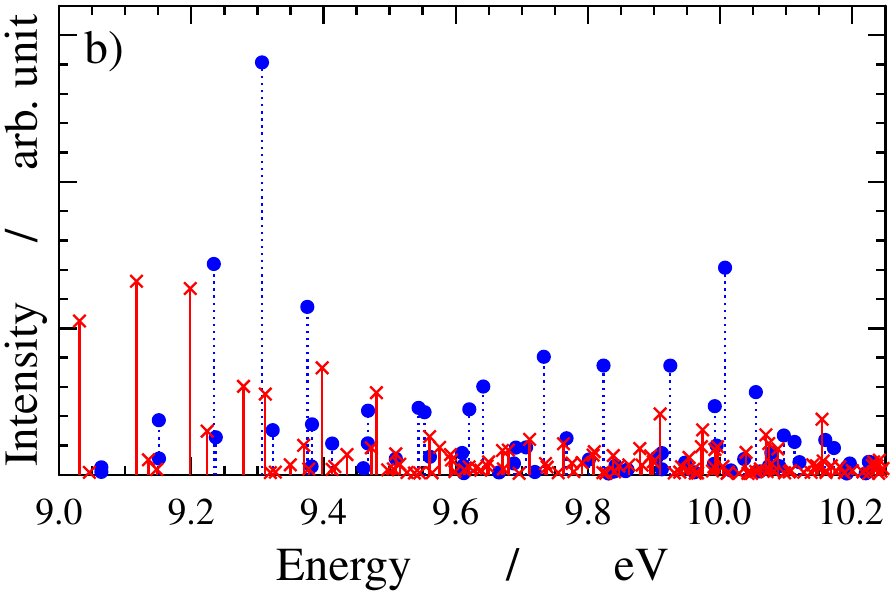}
\caption{\label{fig:4}
        Cavity-free (blue, $\bullet$) and cavity (red, x) ionization spectra of butatriene with cavity parameters
        $\omega_\textrm{cav} = 645.02 ~ \textrm{cm}^{-1}$, $g = 0.002 ~ \textrm{au}$ (panel a),
        and $\omega_\textrm{cav} = 3000.0 ~ \textrm{cm}^{-1}$, $g = 0.01 ~ \textrm{au}$ (panel b).
        In this case, constant transition (TDM, along the $x$ axis) and permanent (PDM, along the $y$ axis)
        dipole moment values of $1.0 ~ \textrm{au}$ are added to the 2-mode model.
        The field polarization is chosen as $\mathbf{e} = (1,1,0)/\sqrt{2}$ (both TDM and PDMs).
        Significant cavity-induced static effects can be observed in the entire spectral range.}
\end{figure*}

Contrary to BT$^{+}$, many typical molecules possess nonzero TDM and/or PDM values at the FC point.
In such cases, there is no need for dynamical symmetry breaking to achieve a strong impact of the cavity on the ionization spectrum.
Consequently, the ionization spectrum is expected to show striking cavity-induced effects which one may call static effects,
meaning that no dynamical symmetry breaking is involved. In order to demonstrate the emergence of static effects in the ionization spectrum,
constant TDM (along the $x$ axis) and PDM (along the $y$ axis) values of $1.0 ~ \textrm{au}$ are added artificially to the original 2-mode model of BT$^+$.
Fig. \ref{fig:4} presents cavity ionization spectra for the modified 2-mode model with cavity parameters
$\omega_\textrm{cav} = 645.02 ~ \textrm{cm}^{-1}$ and $g = 0.002 ~ \textrm{au}$ (panel a), and
$\omega_\textrm{cav} = 3000.0 ~ \textrm{cm}^{-1}$ and $g = 0.01 ~ \textrm{au}$ (panel b),
both obtained with $\mathbf{e} = (1,1,0)/\sqrt{2}$ (both TDM and PDMs are included).
As clearly visible in Fig. \ref{fig:4}, cavity-induced effects are more pronounced than in Fig. \ref{fig:3} and
they affect the entire energy range of the ionization spectra, including the adiabatic region.
These observations are remarkable in light of the much lower coupling strengths $g = 0.002 ~ \textrm{au}$ and
$g = 0.01 ~ \textrm{au}$ (used in Fig. \ref{fig:4}) compared to $g = 0.1 ~ \textrm{au}$ (used in Fig. \ref{fig:3}).

We demonstrate the nonadiabatic effects emerging in the cavity by an explicit example. The system of interest is the butatriene cation, BT$^+$, 
which possesses a natural CI between its ground and first excited electronic states near the Franck--Condon (FC) region of the neutral molecule. 
The natural CI yields a well-separated band (``mystery band'') in the cavity-free ionization spectrum. The 2-mode vibronic coupling model treating a 
coupling and a tuning vibrational mode has been shown to reproduce the cavity-free low-energy ionization spectrum with good accuracy.\cite{Cattarius2001}
Surprisingly, these two modes do not couple to the cavity and the ionization spectrum within this 2-mode model does not exhibit any impact of the cavity. 
We show that the cavity activates other modes of the system than these two modes. Including such modes, indeed gives rise to substantial nonadiabatic
effects induced by the cavity. The origin of these effects is discussed and analyzed in detail.

In order to generate cavity-molecule interactions, one needs to break the symmetry of BT$^+$. This symmetry breaking takes place by the dynamics
of the system after ionization and causes noticeable changes primarily in the nonadiabatic region of the ionization spectrum (that is, above the energetic 
position of the natural CI). This impact of the cavity on the nonadiabatic regime is explained by the proximity of the light-induced CI to the natural CI. 
Our example, butatriene, has been chosen because of its high symmetry which allows for a transparent discussion of the emerging effects. It should be 
stressed, however, that strong effects can be expected for molecules of low symmetry.  Here, the molecules possess nonzero transition and/or permanent
dipole moments at the FC point and static cavity-induced effects (involving no dynamical symmetry breaking) are already expected to give rise to enormous
changes in the entire range of the ionization spectrum. Such effects have been demonstrated in this work.

\begin{acknowledgements}
The authors are indebted to NKFIH for funding (Grant No. K146096).
The work performed in Budapest received funding from the HUN-REN Hungarian Research Network.
Financial support by the Deutsche Forschungsgemeinschaft (DFG) (Grant No. CE 10/56-1) is gratefully acknowledged.
\end{acknowledgements}

\bibliography{cavityIonization}

%aipnum4-2.bst 2019-01-14 (MD) hand-edited version of apsrev4-1.bst
\begin{thebibliography}{76}%
\makeatletter
\providecommand \@ifxundefined [1]{%
 \@ifx{#1\undefined}
}%
\providecommand \@ifnum [1]{%
 \ifnum #1\expandafter \@firstoftwo
 \else \expandafter \@secondoftwo
 \fi
}%
\providecommand \@ifx [1]{%
 \ifx #1\expandafter \@firstoftwo
 \else \expandafter \@secondoftwo
 \fi
}%
\providecommand \natexlab [1]{#1}%
\providecommand \enquote  [1]{``#1''}%
\providecommand \bibnamefont  [1]{#1}%
\providecommand \bibfnamefont [1]{#1}%
\providecommand \citenamefont [1]{#1}%
\providecommand \href@noop [0]{\@secondoftwo}%
\providecommand \href [0]{\begingroup \@sanitize@url \@href}%
\providecommand \@href[1]{\@@startlink{#1}\@@href}%
\providecommand \@@href[1]{\endgroup#1\@@endlink}%
\providecommand \@sanitize@url [0]{\catcode `\\12\catcode `\$12\catcode
  `\&12\catcode `\#12\catcode `\^12\catcode `\_12\catcode `\%12\relax}%
\providecommand \@@startlink[1]{}%
\providecommand \@@endlink[0]{}%
\providecommand \url  [0]{\begingroup\@sanitize@url \@url }%
\providecommand \@url [1]{\endgroup\@href {#1}{\urlprefix }}%
\providecommand \urlprefix  [0]{URL }%
\providecommand \Eprint [0]{\href }%
\providecommand \doibase [0]{https://doi.org/}%
\providecommand \selectlanguage [0]{\@gobble}%
\providecommand \bibinfo  [0]{\@secondoftwo}%
\providecommand \bibfield  [0]{\@secondoftwo}%
\providecommand \translation [1]{[#1]}%
\providecommand \BibitemOpen [0]{}%
\providecommand \bibitemStop [0]{}%
\providecommand \bibitemNoStop [0]{.\EOS\space}%
\providecommand \EOS [0]{\spacefactor3000\relax}%
\providecommand \BibitemShut  [1]{\csname bibitem#1\endcsname}%
\let\auto@bib@innerbib\@empty
%</preamble>
\bibitem [{\citenamefont {Hutchison}\ \emph {et~al.}(2012)\citenamefont
  {Hutchison}, \citenamefont {Schwartz}, \citenamefont {Genet}, \citenamefont
  {Devaux},\ and\ \citenamefont {Ebbesen}}]{Hutchison2012}%
  \BibitemOpen
  \bibfield  {author} {\bibinfo {author} {\bibfnamefont {J.~A.}\ \bibnamefont
  {Hutchison}}, \bibinfo {author} {\bibfnamefont {T.}~\bibnamefont {Schwartz}},
  \bibinfo {author} {\bibfnamefont {C.}~\bibnamefont {Genet}}, \bibinfo
  {author} {\bibfnamefont {E.}~\bibnamefont {Devaux}},\ and\ \bibinfo {author}
  {\bibfnamefont {T.~W.}\ \bibnamefont {Ebbesen}},\ }\bibfield  {title}
  {\enquote {\bibinfo {title} {Modifying chemical landscapes by coupling to
  vacuum fields},}\ }\href {https://doi.org/10.1002/anie.201107033} {\bibfield
  {journal} {\bibinfo  {journal} {Angew. Chem. Int. Ed.}\ }\textbf {\bibinfo
  {volume} {51}},\ \bibinfo {pages} {1592--1596} (\bibinfo {year}
  {2012})}\BibitemShut {NoStop}%
\bibitem [{\citenamefont {Ebbesen}(2016)}]{Ebbesen20162403}%
  \BibitemOpen
  \bibfield  {author} {\bibinfo {author} {\bibfnamefont {T.}~\bibnamefont
  {Ebbesen}},\ }\bibfield  {title} {\enquote {\bibinfo {title} {Hybrid
  light-matter states in a molecular and material science perspective},}\
  }\href {https://doi.org/10.1021/acs.accounts.6b00295} {\bibfield  {journal}
  {\bibinfo  {journal} {Acc. Chem. Res.}\ }\textbf {\bibinfo {volume} {49}},\
  \bibinfo {pages} {2403--2412} (\bibinfo {year} {2016})}\BibitemShut {NoStop}%
\bibitem [{\citenamefont {Chikkaraddy}\ \emph {et~al.}(2016)\citenamefont
  {Chikkaraddy}, \citenamefont {De~Nijs}, \citenamefont {Benz}, \citenamefont
  {Barrow}, \citenamefont {Scherman}, \citenamefont {Rosta}, \citenamefont
  {Demetriadou}, \citenamefont {Fox}, \citenamefont {Hess},\ and\ \citenamefont
  {Baumberg}}]{Chikkaraddy2016127}%
  \BibitemOpen
  \bibfield  {author} {\bibinfo {author} {\bibfnamefont {R.}~\bibnamefont
  {Chikkaraddy}}, \bibinfo {author} {\bibfnamefont {B.}~\bibnamefont
  {De~Nijs}}, \bibinfo {author} {\bibfnamefont {F.}~\bibnamefont {Benz}},
  \bibinfo {author} {\bibfnamefont {S.}~\bibnamefont {Barrow}}, \bibinfo
  {author} {\bibfnamefont {O.}~\bibnamefont {Scherman}}, \bibinfo {author}
  {\bibfnamefont {E.}~\bibnamefont {Rosta}}, \bibinfo {author} {\bibfnamefont
  {A.}~\bibnamefont {Demetriadou}}, \bibinfo {author} {\bibfnamefont
  {P.}~\bibnamefont {Fox}}, \bibinfo {author} {\bibfnamefont {O.}~\bibnamefont
  {Hess}},\ and\ \bibinfo {author} {\bibfnamefont {J.}~\bibnamefont
  {Baumberg}},\ }\bibfield  {title} {\enquote {\bibinfo {title}
  {Single-molecule strong coupling at room temperature in plasmonic
  nanocavities},}\ }\href {https://doi.org/10.1038/nature17974} {\bibfield
  {journal} {\bibinfo  {journal} {Nature}\ }\textbf {\bibinfo {volume} {535}},\
  \bibinfo {pages} {127--130} (\bibinfo {year} {2016})}\BibitemShut {NoStop}%
\bibitem [{\citenamefont {Kowalewski}, \citenamefont {Bennett},\ and\
  \citenamefont {Mukamel}(2016)}]{Kowalewski20162050}%
  \BibitemOpen
  \bibfield  {author} {\bibinfo {author} {\bibfnamefont {M.}~\bibnamefont
  {Kowalewski}}, \bibinfo {author} {\bibfnamefont {K.}~\bibnamefont
  {Bennett}},\ and\ \bibinfo {author} {\bibfnamefont {S.}~\bibnamefont
  {Mukamel}},\ }\bibfield  {title} {\enquote {\bibinfo {title} {Cavity
  femtochemistry: Manipulating nonadiabatic dynamics at avoided crossings},}\
  }\href {https://doi.org/10.1021/acs.jpclett.6b00864} {\bibfield  {journal}
  {\bibinfo  {journal} {J. Phys. Chem. Lett.}\ }\textbf {\bibinfo {volume}
  {7}},\ \bibinfo {pages} {2050--2054} (\bibinfo {year} {2016})}\BibitemShut
  {NoStop}%
\bibitem [{\citenamefont {Flick}\ \emph {et~al.}(2017)\citenamefont {Flick},
  \citenamefont {Ruggenthaler}, \citenamefont {Appel},\ and\ \citenamefont
  {Rubio}}]{Flick20173026}%
  \BibitemOpen
  \bibfield  {author} {\bibinfo {author} {\bibfnamefont {J.}~\bibnamefont
  {Flick}}, \bibinfo {author} {\bibfnamefont {M.}~\bibnamefont {Ruggenthaler}},
  \bibinfo {author} {\bibfnamefont {H.}~\bibnamefont {Appel}},\ and\ \bibinfo
  {author} {\bibfnamefont {A.}~\bibnamefont {Rubio}},\ }\bibfield  {title}
  {\enquote {\bibinfo {title} {Atoms and molecules in cavities, from weak to
  strong coupling in quantum-electrodynamics ({QED}) chemistry},}\ }\href
  {https://doi.org/10.1073/pnas.1615509114} {\bibfield  {journal} {\bibinfo
  {journal} {Proc. Natl. Acad. Sci. U.S.A.}\ }\textbf {\bibinfo {volume}
  {114}},\ \bibinfo {pages} {3026--3034} (\bibinfo {year} {2017})}\BibitemShut
  {NoStop}%
\bibitem [{\citenamefont {Feist}, \citenamefont {Galego},\ and\ \citenamefont
  {Garcia-Vidal}(2018)}]{Feist2018205}%
  \BibitemOpen
  \bibfield  {author} {\bibinfo {author} {\bibfnamefont {J.}~\bibnamefont
  {Feist}}, \bibinfo {author} {\bibfnamefont {J.}~\bibnamefont {Galego}},\ and\
  \bibinfo {author} {\bibfnamefont {F.}~\bibnamefont {Garcia-Vidal}},\
  }\bibfield  {title} {\enquote {\bibinfo {title} {Polaritonic chemistry with
  organic molecules},}\ }\href {https://doi.org/10.1021/acsphotonics.7b00680}
  {\bibfield  {journal} {\bibinfo  {journal} {ACS Photonics}\ }\textbf
  {\bibinfo {volume} {5}},\ \bibinfo {pages} {205--216} (\bibinfo {year}
  {2018})}\BibitemShut {NoStop}%
\bibitem [{\citenamefont {Fregoni}\ \emph {et~al.}(2018)\citenamefont
  {Fregoni}, \citenamefont {Granucci}, \citenamefont {Coccia}, \citenamefont
  {Persico},\ and\ \citenamefont {Corni}}]{Fregoni2018}%
  \BibitemOpen
  \bibfield  {author} {\bibinfo {author} {\bibfnamefont {J.}~\bibnamefont
  {Fregoni}}, \bibinfo {author} {\bibfnamefont {G.}~\bibnamefont {Granucci}},
  \bibinfo {author} {\bibfnamefont {E.}~\bibnamefont {Coccia}}, \bibinfo
  {author} {\bibfnamefont {M.}~\bibnamefont {Persico}},\ and\ \bibinfo {author}
  {\bibfnamefont {S.}~\bibnamefont {Corni}},\ }\bibfield  {title} {\enquote
  {\bibinfo {title} {Manipulating azobenzene photoisomerization through strong
  light-molecule coupling},}\ }\href
  {https://doi.org/10.1038/s41467-018-06971-y} {\bibfield  {journal} {\bibinfo
  {journal} {Nat. Commun.}\ }\textbf {\bibinfo {volume} {9}},\ \bibinfo {pages}
  {4688} (\bibinfo {year} {2018})}\BibitemShut {NoStop}%
\bibitem [{\citenamefont {Ribeiro}\ \emph {et~al.}(2018)\citenamefont
  {Ribeiro}, \citenamefont {Mart\'inez-Mart\'inez}, \citenamefont {Du},
  \citenamefont {Campos-Gonzalez-Angulo},\ and\ \citenamefont
  {Yuen-Zhou}}]{Ribeiro20186325}%
  \BibitemOpen
  \bibfield  {author} {\bibinfo {author} {\bibfnamefont {R.}~\bibnamefont
  {Ribeiro}}, \bibinfo {author} {\bibfnamefont {L.}~\bibnamefont
  {Mart\'inez-Mart\'inez}}, \bibinfo {author} {\bibfnamefont {M.}~\bibnamefont
  {Du}}, \bibinfo {author} {\bibfnamefont {J.}~\bibnamefont
  {Campos-Gonzalez-Angulo}},\ and\ \bibinfo {author} {\bibfnamefont
  {J.}~\bibnamefont {Yuen-Zhou}},\ }\bibfield  {title} {\enquote {\bibinfo
  {title} {Polariton chemistry: Controlling molecular dynamics with optical
  cavities},}\ }\href {https://doi.org/10.1039/c8sc01043a} {\bibfield
  {journal} {\bibinfo  {journal} {Chem. Sci.}\ }\textbf {\bibinfo {volume}
  {9}},\ \bibinfo {pages} {6325--6339} (\bibinfo {year} {2018})}\BibitemShut
  {NoStop}%
\bibitem [{\citenamefont {Szidarovszky}\ \emph {et~al.}(2018)\citenamefont
  {Szidarovszky}, \citenamefont {Hal\'asz}, \citenamefont {Cs\'asz\'ar},
  \citenamefont {Cederbaum},\ and\ \citenamefont
  {Vib\'ok}}]{Szidarovszky20186215}%
  \BibitemOpen
  \bibfield  {author} {\bibinfo {author} {\bibfnamefont {T.}~\bibnamefont
  {Szidarovszky}}, \bibinfo {author} {\bibfnamefont {G.}~\bibnamefont
  {Hal\'asz}}, \bibinfo {author} {\bibfnamefont {A.}~\bibnamefont
  {Cs\'asz\'ar}}, \bibinfo {author} {\bibfnamefont {L.}~\bibnamefont
  {Cederbaum}},\ and\ \bibinfo {author} {\bibfnamefont {{\'A}.}~\bibnamefont
  {Vib\'ok}},\ }\bibfield  {title} {\enquote {\bibinfo {title} {Conical
  intersections induced by quantum light: Field-dressed spectra from the weak
  to the ultrastrong coupling regimes},}\ }\href
  {https://doi.org/10.1021/acs.jpclett.8b02609} {\bibfield  {journal} {\bibinfo
   {journal} {J. Phys. Chem. Lett.}\ }\textbf {\bibinfo {volume} {9}},\
  \bibinfo {pages} {6215--6223} (\bibinfo {year} {2018})}\BibitemShut {NoStop}%
\bibitem [{\citenamefont {Ruggenthaler}\ \emph {et~al.}(2018)\citenamefont
  {Ruggenthaler}, \citenamefont {Tancogne-Dejean}, \citenamefont {Flick},
  \citenamefont {Appel},\ and\ \citenamefont {Rubio}}]{Ruggenthaler2018}%
  \BibitemOpen
  \bibfield  {author} {\bibinfo {author} {\bibfnamefont {M.}~\bibnamefont
  {Ruggenthaler}}, \bibinfo {author} {\bibfnamefont {N.}~\bibnamefont
  {Tancogne-Dejean}}, \bibinfo {author} {\bibfnamefont {J.}~\bibnamefont
  {Flick}}, \bibinfo {author} {\bibfnamefont {H.}~\bibnamefont {Appel}},\ and\
  \bibinfo {author} {\bibfnamefont {A.}~\bibnamefont {Rubio}},\ }\bibfield
  {title} {\enquote {\bibinfo {title} {From a quantum-electrodynamical
  light-matter description to novel spectroscopies},}\ }\href
  {https://doi.org/10.1038/s41570-018-0118} {\bibfield  {journal} {\bibinfo
  {journal} {Nat. Rev. Chem.}\ }\textbf {\bibinfo {volume} {2}},\ \bibinfo
  {pages} {0118} (\bibinfo {year} {2018})}\BibitemShut {NoStop}%
\bibitem [{\citenamefont {Vendrell}(2018)}]{Vendrell2018}%
  \BibitemOpen
  \bibfield  {author} {\bibinfo {author} {\bibfnamefont {O.}~\bibnamefont
  {Vendrell}},\ }\bibfield  {title} {\enquote {\bibinfo {title} {Collective
  {Jahn-Teller} interactions through light-matter coupling in a cavity},}\
  }\href {https://doi.org/10.1103/PhysRevLett.121.253001} {\bibfield  {journal}
  {\bibinfo  {journal} {Phys. Rev. Lett.}\ }\textbf {\bibinfo {volume} {121}},\
  \bibinfo {pages} {253001} (\bibinfo {year} {2018})}\BibitemShut {NoStop}%
\bibitem [{\citenamefont {Csehi}\ \emph
  {et~al.}(2019{\natexlab{a}})\citenamefont {Csehi}, \citenamefont {Vib\'ok},
  \citenamefont {Hal\'asz},\ and\ \citenamefont {Kowalewski}}]{Csehi2019}%
  \BibitemOpen
  \bibfield  {author} {\bibinfo {author} {\bibfnamefont {A.}~\bibnamefont
  {Csehi}}, \bibinfo {author} {\bibfnamefont {{\'A}.}~\bibnamefont {Vib\'ok}},
  \bibinfo {author} {\bibfnamefont {G.}~\bibnamefont {Hal\'asz}},\ and\
  \bibinfo {author} {\bibfnamefont {M.}~\bibnamefont {Kowalewski}},\ }\bibfield
   {title} {\enquote {\bibinfo {title} {Quantum control with quantum light of
  molecular nonadiabaticity},}\ }\href
  {https://doi.org/10.1103/PhysRevA.100.053421} {\bibfield  {journal} {\bibinfo
   {journal} {Phys. Rev. A}\ }\textbf {\bibinfo {volume} {100}},\ \bibinfo
  {pages} {053421} (\bibinfo {year} {2019}{\natexlab{a}})}\BibitemShut
  {NoStop}%
\bibitem [{\citenamefont {Csehi}\ \emph
  {et~al.}(2019{\natexlab{b}})\citenamefont {Csehi}, \citenamefont
  {Kowalewski}, \citenamefont {Hal\'asz},\ and\ \citenamefont
  {Vib\'ok}}]{Csehi2019a}%
  \BibitemOpen
  \bibfield  {author} {\bibinfo {author} {\bibfnamefont {A.}~\bibnamefont
  {Csehi}}, \bibinfo {author} {\bibfnamefont {M.}~\bibnamefont {Kowalewski}},
  \bibinfo {author} {\bibfnamefont {G.}~\bibnamefont {Hal\'asz}},\ and\
  \bibinfo {author} {\bibfnamefont {{\'A}.}~\bibnamefont {Vib\'ok}},\
  }\bibfield  {title} {\enquote {\bibinfo {title} {Ultrafast dynamics in the
  vicinity of quantum light-induced conical intersections},}\ }\href
  {https://doi.org/10.1088/1367-2630/ab3fcc} {\bibfield  {journal} {\bibinfo
  {journal} {New J. Phys.}\ }\textbf {\bibinfo {volume} {21}},\ \bibinfo
  {pages} {093040} (\bibinfo {year} {2019}{\natexlab{b}})}\BibitemShut
  {NoStop}%
\bibitem [{\citenamefont {Reitz}, \citenamefont {Sommer},\ and\ \citenamefont
  {Genes}(2019)}]{Reitz2019}%
  \BibitemOpen
  \bibfield  {author} {\bibinfo {author} {\bibfnamefont {M.}~\bibnamefont
  {Reitz}}, \bibinfo {author} {\bibfnamefont {C.}~\bibnamefont {Sommer}},\ and\
  \bibinfo {author} {\bibfnamefont {C.}~\bibnamefont {Genes}},\ }\bibfield
  {title} {\enquote {\bibinfo {title} {Langevin approach to quantum optics with
  molecules},}\ }\href {https://doi.org/10.1103/PhysRevLett.122.203602}
  {\bibfield  {journal} {\bibinfo  {journal} {Phys. Rev. Lett.}\ }\textbf
  {\bibinfo {volume} {122}},\ \bibinfo {pages} {203602} (\bibinfo {year}
  {2019})}\BibitemShut {NoStop}%
\bibitem [{\citenamefont {Ulusoy}, \citenamefont {Gomez},\ and\ \citenamefont
  {Vendrell}(2019)}]{Ulusoy20198832}%
  \BibitemOpen
  \bibfield  {author} {\bibinfo {author} {\bibfnamefont {I.}~\bibnamefont
  {Ulusoy}}, \bibinfo {author} {\bibfnamefont {J.}~\bibnamefont {Gomez}},\ and\
  \bibinfo {author} {\bibfnamefont {O.}~\bibnamefont {Vendrell}},\ }\bibfield
  {title} {\enquote {\bibinfo {title} {Modifying the nonradiative decay
  dynamics through conical intersections via collective coupling to a cavity
  mode},}\ }\href {https://doi.org/10.1021/acs.jpca.9b07404} {\bibfield
  {journal} {\bibinfo  {journal} {J. Phys. Chem. A}\ }\textbf {\bibinfo
  {volume} {123}},\ \bibinfo {pages} {8832--8844} (\bibinfo {year}
  {2019})}\BibitemShut {NoStop}%
\bibitem [{\citenamefont {Mandal}\ and\ \citenamefont
  {Huo}(2019)}]{Mandal2019}%
  \BibitemOpen
  \bibfield  {author} {\bibinfo {author} {\bibfnamefont {A.}~\bibnamefont
  {Mandal}}\ and\ \bibinfo {author} {\bibfnamefont {P.}~\bibnamefont {Huo}},\
  }\bibfield  {title} {\enquote {\bibinfo {title} {Investigating new
  reactivities enabled by polariton photochemistry},}\ }\href
  {https://doi.org/10.1021/acs.jpclett.9b01599} {\bibfield  {journal} {\bibinfo
   {journal} {J. Phys. Chem. Lett.}\ }\textbf {\bibinfo {volume} {10}},\
  \bibinfo {pages} {5519--5529} (\bibinfo {year} {2019})}\BibitemShut {NoStop}%
\bibitem [{\citenamefont {Triana}\ and\ \citenamefont
  {Sanz-Vicario}(2019)}]{Triana2019}%
  \BibitemOpen
  \bibfield  {author} {\bibinfo {author} {\bibfnamefont {J.}~\bibnamefont
  {Triana}}\ and\ \bibinfo {author} {\bibfnamefont {J.}~\bibnamefont
  {Sanz-Vicario}},\ }\bibfield  {title} {\enquote {\bibinfo {title} {Revealing
  the presence of potential crossings in diatomics induced by quantum cavity
  radiation},}\ }\href {https://doi.org/10.1103/PhysRevLett.122.063603}
  {\bibfield  {journal} {\bibinfo  {journal} {Phys. Rev. Lett.}\ }\textbf
  {\bibinfo {volume} {122}},\ \bibinfo {pages} {063603} (\bibinfo {year}
  {2019})}\BibitemShut {NoStop}%
\bibitem [{\citenamefont {Yuen-Zhou}\ and\ \citenamefont
  {Menon}(2019)}]{19YuMe}%
  \BibitemOpen
  \bibfield  {author} {\bibinfo {author} {\bibfnamefont {J.}~\bibnamefont
  {Yuen-Zhou}}\ and\ \bibinfo {author} {\bibfnamefont {V.~M.}\ \bibnamefont
  {Menon}},\ }\bibfield  {title} {\enquote {\bibinfo {title} {Polariton
  chemistry: Thinking inside the (photon) box},}\ }\href
  {https://doi.org/10.1073/pnas.1900795116} {\bibfield  {journal} {\bibinfo
  {journal} {Proc. Natl. Acad. Sci. U.S.A.}\ }\textbf {\bibinfo {volume}
  {116}},\ \bibinfo {pages} {5214--5216} (\bibinfo {year} {2019})}\BibitemShut
  {NoStop}%
\bibitem [{\citenamefont {Davidsson}\ and\ \citenamefont
  {Kowalewski}(2020)}]{Davidsson2020234304}%
  \BibitemOpen
  \bibfield  {author} {\bibinfo {author} {\bibfnamefont {E.}~\bibnamefont
  {Davidsson}}\ and\ \bibinfo {author} {\bibfnamefont {M.}~\bibnamefont
  {Kowalewski}},\ }\bibfield  {title} {\enquote {\bibinfo {title} {Simulating
  photodissociation reactions in bad cavities with the {Lindblad} equation},}\
  }\href {https://doi.org/10.1063/5.0033773} {\bibfield  {journal} {\bibinfo
  {journal} {J. Chem. Phys.}\ }\textbf {\bibinfo {volume} {153}},\ \bibinfo
  {pages} {234304} (\bibinfo {year} {2020})}\BibitemShut {NoStop}%
\bibitem [{\citenamefont {F\'abri}\ \emph {et~al.}(2020)\citenamefont
  {F\'abri}, \citenamefont {Lasorne}, \citenamefont {Hal\'asz}, \citenamefont
  {Cederbaum},\ and\ \citenamefont {Vib\'ok}}]{Fabri2020234302}%
  \BibitemOpen
  \bibfield  {author} {\bibinfo {author} {\bibfnamefont {C.}~\bibnamefont
  {F\'abri}}, \bibinfo {author} {\bibfnamefont {B.}~\bibnamefont {Lasorne}},
  \bibinfo {author} {\bibfnamefont {G.}~\bibnamefont {Hal\'asz}}, \bibinfo
  {author} {\bibfnamefont {L.}~\bibnamefont {Cederbaum}},\ and\ \bibinfo
  {author} {\bibfnamefont {{\'A}.}~\bibnamefont {Vib\'ok}},\ }\bibfield
  {title} {\enquote {\bibinfo {title} {Quantum light-induced nonadiabatic
  phenomena in the absorption spectrum of formaldehyde: Full- and
  reduced-dimensionality studies},}\ }\href {https://doi.org/10.1063/5.0035870}
  {\bibfield  {journal} {\bibinfo  {journal} {J. Chem. Phys.}\ }\textbf
  {\bibinfo {volume} {153}},\ \bibinfo {pages} {234302} (\bibinfo {year}
  {2020})}\BibitemShut {NoStop}%
\bibitem [{\citenamefont {Gu}\ and\ \citenamefont
  {Mukamel}(2020{\natexlab{a}})}]{Gu20201290}%
  \BibitemOpen
  \bibfield  {author} {\bibinfo {author} {\bibfnamefont {B.}~\bibnamefont
  {Gu}}\ and\ \bibinfo {author} {\bibfnamefont {S.}~\bibnamefont {Mukamel}},\
  }\bibfield  {title} {\enquote {\bibinfo {title} {Manipulating nonadiabatic
  conical intersection dynamics by optical cavities},}\ }\href
  {https://doi.org/10.1039/c9sc04992d} {\bibfield  {journal} {\bibinfo
  {journal} {Chem. Sci.}\ }\textbf {\bibinfo {volume} {11}},\ \bibinfo {pages}
  {1290--1298} (\bibinfo {year} {2020}{\natexlab{a}})}\BibitemShut {NoStop}%
\bibitem [{\citenamefont {Gu}\ and\ \citenamefont
  {Mukamel}(2020{\natexlab{b}})}]{Gu2020a}%
  \BibitemOpen
  \bibfield  {author} {\bibinfo {author} {\bibfnamefont {B.}~\bibnamefont
  {Gu}}\ and\ \bibinfo {author} {\bibfnamefont {S.}~\bibnamefont {Mukamel}},\
  }\bibfield  {title} {\enquote {\bibinfo {title} {Cooperative conical
  intersection dynamics of two pyrazine molecules in an optical cavity},}\
  }\href {https://doi.org/10.1021/acs.jpclett.0c00381} {\bibfield  {journal}
  {\bibinfo  {journal} {J. Phys. Chem. Lett.}\ }\textbf {\bibinfo {volume}
  {11}},\ \bibinfo {pages} {5555--5562} (\bibinfo {year}
  {2020}{\natexlab{b}})}\BibitemShut {NoStop}%
\bibitem [{\citenamefont {Silva}\ \emph {et~al.}(2020)\citenamefont {Silva},
  \citenamefont {Pino}, \citenamefont {Garc\'ia-Vidal},\ and\ \citenamefont
  {Feist}}]{Silva2020}%
  \BibitemOpen
  \bibfield  {author} {\bibinfo {author} {\bibfnamefont {R.}~\bibnamefont
  {Silva}}, \bibinfo {author} {\bibfnamefont {J.}~\bibnamefont {Pino}},
  \bibinfo {author} {\bibfnamefont {F.}~\bibnamefont {Garc\'ia-Vidal}},\ and\
  \bibinfo {author} {\bibfnamefont {J.}~\bibnamefont {Feist}},\ }\bibfield
  {title} {\enquote {\bibinfo {title} {Polaritonic molecular clock for
  all-optical ultrafast imaging of wavepacket dynamics without probe pulses},}\
  }\href {https://doi.org/10.1038/s41467-020-15196-x} {\bibfield  {journal}
  {\bibinfo  {journal} {Nat. Commun.}\ }\textbf {\bibinfo {volume} {11}},\
  \bibinfo {pages} {1423} (\bibinfo {year} {2020})}\BibitemShut {NoStop}%
\bibitem [{\citenamefont {Felicetti}\ \emph {et~al.}(2020)\citenamefont
  {Felicetti}, \citenamefont {Fregoni}, \citenamefont {Schnappinger},
  \citenamefont {Reiter}, \citenamefont {De~Vivie-Riedle},\ and\ \citenamefont
  {Feist}}]{Felicetti20208810}%
  \BibitemOpen
  \bibfield  {author} {\bibinfo {author} {\bibfnamefont {S.}~\bibnamefont
  {Felicetti}}, \bibinfo {author} {\bibfnamefont {J.}~\bibnamefont {Fregoni}},
  \bibinfo {author} {\bibfnamefont {T.}~\bibnamefont {Schnappinger}}, \bibinfo
  {author} {\bibfnamefont {S.}~\bibnamefont {Reiter}}, \bibinfo {author}
  {\bibfnamefont {R.}~\bibnamefont {De~Vivie-Riedle}},\ and\ \bibinfo {author}
  {\bibfnamefont {J.}~\bibnamefont {Feist}},\ }\bibfield  {title} {\enquote
  {\bibinfo {title} {Photoprotecting uracil by coupling with lossy
  nanocavities},}\ }\href {https://doi.org/10.1021/acs.jpclett.0c02236}
  {\bibfield  {journal} {\bibinfo  {journal} {J. Phys. Chem. Lett.}\ }\textbf
  {\bibinfo {volume} {11}},\ \bibinfo {pages} {8810--8818} (\bibinfo {year}
  {2020})}\BibitemShut {NoStop}%
\bibitem [{\citenamefont {Fregoni}\ \emph {et~al.}(2020)\citenamefont
  {Fregoni}, \citenamefont {Granucci}, \citenamefont {Persico},\ and\
  \citenamefont {Corni}}]{Fregoni2020}%
  \BibitemOpen
  \bibfield  {author} {\bibinfo {author} {\bibfnamefont {J.}~\bibnamefont
  {Fregoni}}, \bibinfo {author} {\bibfnamefont {G.}~\bibnamefont {Granucci}},
  \bibinfo {author} {\bibfnamefont {M.}~\bibnamefont {Persico}},\ and\ \bibinfo
  {author} {\bibfnamefont {S.}~\bibnamefont {Corni}},\ }\bibfield  {title}
  {\enquote {\bibinfo {title} {Strong coupling with light enhances the
  photoisomerization quantum yield of azobenzene},}\ }\href
  {https://doi.org/10.1016/j.chempr.2019.11.001} {\bibfield  {journal}
  {\bibinfo  {journal} {Chem}\ }\textbf {\bibinfo {volume} {6}},\ \bibinfo
  {pages} {250--265} (\bibinfo {year} {2020})}\BibitemShut {NoStop}%
\bibitem [{\citenamefont {Mandal}, \citenamefont {Krauss},\ and\ \citenamefont
  {Huo}(2020)}]{Mandal2020}%
  \BibitemOpen
  \bibfield  {author} {\bibinfo {author} {\bibfnamefont {A.}~\bibnamefont
  {Mandal}}, \bibinfo {author} {\bibfnamefont {T.~D.}\ \bibnamefont {Krauss}},\
  and\ \bibinfo {author} {\bibfnamefont {P.}~\bibnamefont {Huo}},\ }\bibfield
  {title} {\enquote {\bibinfo {title} {Polariton-mediated electron transfer via
  cavity quantum electrodynamics},}\ }\href
  {https://doi.org/10.1021/acs.jpcb.0c03227} {\bibfield  {journal} {\bibinfo
  {journal} {J. Phys. Chem. B}\ }\textbf {\bibinfo {volume} {124}},\ \bibinfo
  {pages} {6321--6340} (\bibinfo {year} {2020})}\BibitemShut {NoStop}%
\bibitem [{\citenamefont {Polak}\ \emph {et~al.}(2020)\citenamefont {Polak},
  \citenamefont {Jayaprakash}, \citenamefont {Lyons}, \citenamefont
  {Mart\'inez-Mart\'inez}, \citenamefont {Leventis}, \citenamefont {Fallon},
  \citenamefont {Coulthard}, \citenamefont {Bossanyi}, \citenamefont
  {Georgiou}, \citenamefont {Petty{,}~II}, \citenamefont {Anthony},
  \citenamefont {Bronstein}, \citenamefont {Yuen-Zhou}, \citenamefont
  {Tartakovskii}, \citenamefont {Clark},\ and\ \citenamefont
  {Musser}}]{20PoJaLy}%
  \BibitemOpen
  \bibfield  {author} {\bibinfo {author} {\bibfnamefont {D.}~\bibnamefont
  {Polak}}, \bibinfo {author} {\bibfnamefont {R.}~\bibnamefont {Jayaprakash}},
  \bibinfo {author} {\bibfnamefont {T.~P.}\ \bibnamefont {Lyons}}, \bibinfo
  {author} {\bibfnamefont {L.~{\'A}.}\ \bibnamefont {Mart\'inez-Mart\'inez}},
  \bibinfo {author} {\bibfnamefont {A.}~\bibnamefont {Leventis}}, \bibinfo
  {author} {\bibfnamefont {K.~J.}\ \bibnamefont {Fallon}}, \bibinfo {author}
  {\bibfnamefont {H.}~\bibnamefont {Coulthard}}, \bibinfo {author}
  {\bibfnamefont {D.~G.}\ \bibnamefont {Bossanyi}}, \bibinfo {author}
  {\bibfnamefont {K.}~\bibnamefont {Georgiou}}, \bibinfo {author}
  {\bibfnamefont {A.~J.}\ \bibnamefont {Petty{,}~II}}, \bibinfo {author}
  {\bibfnamefont {J.}~\bibnamefont {Anthony}}, \bibinfo {author} {\bibfnamefont
  {H.}~\bibnamefont {Bronstein}}, \bibinfo {author} {\bibfnamefont
  {J.}~\bibnamefont {Yuen-Zhou}}, \bibinfo {author} {\bibfnamefont {A.~I.}\
  \bibnamefont {Tartakovskii}}, \bibinfo {author} {\bibfnamefont
  {J.}~\bibnamefont {Clark}},\ and\ \bibinfo {author} {\bibfnamefont {A.~J.}\
  \bibnamefont {Musser}},\ }\bibfield  {title} {\enquote {\bibinfo {title}
  {Manipulating molecules with strong coupling: Harvesting triplet excitons in
  organic exciton microcavities},}\ }\href {https://doi.org/10.1039/C9SC04950A}
  {\bibfield  {journal} {\bibinfo  {journal} {Chem. Sci.}\ }\textbf {\bibinfo
  {volume} {11}},\ \bibinfo {pages} {343--354} (\bibinfo {year}
  {2020})}\BibitemShut {NoStop}%
\bibitem [{\citenamefont {F\'abri}\ \emph {et~al.}(2021)\citenamefont
  {F\'abri}, \citenamefont {Hal\'asz}, \citenamefont {Cederbaum},\ and\
  \citenamefont {Vib\'ok}}]{Fabri2021a}%
  \BibitemOpen
  \bibfield  {author} {\bibinfo {author} {\bibfnamefont {C.}~\bibnamefont
  {F\'abri}}, \bibinfo {author} {\bibfnamefont {G.}~\bibnamefont {Hal\'asz}},
  \bibinfo {author} {\bibfnamefont {L.}~\bibnamefont {Cederbaum}},\ and\
  \bibinfo {author} {\bibfnamefont {{\'A}.}~\bibnamefont {Vib\'ok}},\
  }\bibfield  {title} {\enquote {\bibinfo {title} {{Born-Oppenheimer}
  approximation in optical cavities: From success to breakdown},}\ }\href
  {https://doi.org/10.1039/d0sc05164k} {\bibfield  {journal} {\bibinfo
  {journal} {Chem. Sci.}\ }\textbf {\bibinfo {volume} {12}},\ \bibinfo {pages}
  {1251--1258} (\bibinfo {year} {2021})}\BibitemShut {NoStop}%
\bibitem [{\citenamefont {Garcia-Vidal}, \citenamefont {Ciuti},\ and\
  \citenamefont {Ebbesen}(2021)}]{Garcia-Vidal2021}%
  \BibitemOpen
  \bibfield  {author} {\bibinfo {author} {\bibfnamefont {F.~J.}\ \bibnamefont
  {Garcia-Vidal}}, \bibinfo {author} {\bibfnamefont {C.}~\bibnamefont
  {Ciuti}},\ and\ \bibinfo {author} {\bibfnamefont {T.~W.}\ \bibnamefont
  {Ebbesen}},\ }\bibfield  {title} {\enquote {\bibinfo {title} {Manipulating
  matter by strong coupling to vacuum fields},}\ }\href
  {https://doi.org/10.1126/science.abd0336} {\bibfield  {journal} {\bibinfo
  {journal} {Science}\ }\textbf {\bibinfo {volume} {373}},\ \bibinfo {pages}
  {eabd0336} (\bibinfo {year} {2021})}\BibitemShut {NoStop}%
\bibitem [{\citenamefont {Huang}\ \emph {et~al.}(2021)\citenamefont {Huang},
  \citenamefont {Ahrens}, \citenamefont {Beutel},\ and\ \citenamefont
  {Varga}}]{Huang2021}%
  \BibitemOpen
  \bibfield  {author} {\bibinfo {author} {\bibfnamefont {C.}~\bibnamefont
  {Huang}}, \bibinfo {author} {\bibfnamefont {A.}~\bibnamefont {Ahrens}},
  \bibinfo {author} {\bibfnamefont {M.}~\bibnamefont {Beutel}},\ and\ \bibinfo
  {author} {\bibfnamefont {K.}~\bibnamefont {Varga}},\ }\bibfield  {title}
  {\enquote {\bibinfo {title} {Two electrons in harmonic confinement coupled to
  light in a cavity},}\ }\href {https://doi.org/10.1103/PhysRevB.104.165147}
  {\bibfield  {journal} {\bibinfo  {journal} {Phys. Rev. B}\ }\textbf {\bibinfo
  {volume} {104}},\ \bibinfo {pages} {165147} (\bibinfo {year}
  {2021})}\BibitemShut {NoStop}%
\bibitem [{\citenamefont {Szidarovszky}\ \emph {et~al.}(2021)\citenamefont
  {Szidarovszky}, \citenamefont {Badank\'o}, \citenamefont {Hal\'asz},\ and\
  \citenamefont {Vib\'ok}}]{Szidarovszky2021}%
  \BibitemOpen
  \bibfield  {author} {\bibinfo {author} {\bibfnamefont {T.}~\bibnamefont
  {Szidarovszky}}, \bibinfo {author} {\bibfnamefont {P.}~\bibnamefont
  {Badank\'o}}, \bibinfo {author} {\bibfnamefont {G.~J.}\ \bibnamefont
  {Hal\'asz}},\ and\ \bibinfo {author} {\bibfnamefont {{\'A}.}~\bibnamefont
  {Vib\'ok}},\ }\bibfield  {title} {\enquote {\bibinfo {title} {{Nonadiabatic
  Phenomena in Molecular Vibrational Polaritons}},}\ }\href
  {https://doi.org/10.1063/5.0033338} {\bibfield  {journal} {\bibinfo
  {journal} {J. Chem. Phys.}\ }\textbf {\bibinfo {volume} {154}},\ \bibinfo
  {pages} {064305} (\bibinfo {year} {2021})}\BibitemShut {NoStop}%
\bibitem [{\citenamefont {Cederbaum}\ and\ \citenamefont
  {Kuleff}(2021)}]{Cederbaum2021}%
  \BibitemOpen
  \bibfield  {author} {\bibinfo {author} {\bibfnamefont {L.~S.}\ \bibnamefont
  {Cederbaum}}\ and\ \bibinfo {author} {\bibfnamefont {A.~I.}\ \bibnamefont
  {Kuleff}},\ }\bibfield  {title} {\enquote {\bibinfo {title} {Impact of cavity
  on interatomic coulombic decay},}\ }\href
  {https://doi.org/10.1038/s41467-021-24221-6} {\bibfield  {journal} {\bibinfo
  {journal} {Nat. Commun.}\ }\textbf {\bibinfo {volume} {12}},\ \bibinfo
  {pages} {4083} (\bibinfo {year} {2021})}\BibitemShut {NoStop}%
\bibitem [{\citenamefont {Cederbaum}(2021)}]{Cederbaum2021a}%
  \BibitemOpen
  \bibfield  {author} {\bibinfo {author} {\bibfnamefont {L.~S.}\ \bibnamefont
  {Cederbaum}},\ }\bibfield  {title} {\enquote {\bibinfo {title} {Polaritonic
  states of matter in a rotating cavity},}\ }\href
  {https://doi.org/10.1021/acs.jpclett.1c01570} {\bibfield  {journal} {\bibinfo
   {journal} {J. Phys. Chem. Lett.}\ }\textbf {\bibinfo {volume} {12}},\
  \bibinfo {pages} {6056--6061} (\bibinfo {year} {2021})}\BibitemShut {NoStop}%
\bibitem [{\citenamefont {Sch\"{a}fer}\ \emph {et~al.}(2021)\citenamefont
  {Sch\"{a}fer}, \citenamefont {Buchholz}, \citenamefont {Penz}, \citenamefont
  {Ruggenthaler},\ and\ \citenamefont {Rubio}}]{Schfer2021}%
  \BibitemOpen
  \bibfield  {author} {\bibinfo {author} {\bibfnamefont {C.}~\bibnamefont
  {Sch\"{a}fer}}, \bibinfo {author} {\bibfnamefont {F.}~\bibnamefont
  {Buchholz}}, \bibinfo {author} {\bibfnamefont {M.}~\bibnamefont {Penz}},
  \bibinfo {author} {\bibfnamefont {M.}~\bibnamefont {Ruggenthaler}},\ and\
  \bibinfo {author} {\bibfnamefont {A.}~\bibnamefont {Rubio}},\ }\bibfield
  {title} {\enquote {\bibinfo {title} {Making ab initio {QED} functional(s):
  Nonperturbative and photon-free effective frameworks for strong light-matter
  coupling},}\ }\href {https://doi.org/10.1073/pnas.2110464118} {\bibfield
  {journal} {\bibinfo  {journal} {Proc. Natl. Acad. Sci. U.S.A.}\ }\textbf
  {\bibinfo {volume} {118}},\ \bibinfo {pages} {e2110464118} (\bibinfo {year}
  {2021})}\BibitemShut {NoStop}%
\bibitem [{\citenamefont {Triana}\ and\ \citenamefont
  {Sanz-Vicario}(2021)}]{Triana2021}%
  \BibitemOpen
  \bibfield  {author} {\bibinfo {author} {\bibfnamefont {J.}~\bibnamefont
  {Triana}}\ and\ \bibinfo {author} {\bibfnamefont {J.}~\bibnamefont
  {Sanz-Vicario}},\ }\bibfield  {title} {\enquote {\bibinfo {title} {Polar
  diatomic molecules in optical cavities: Photon scaling, rotational effects,
  and comparison with classical fields},}\ }\href
  {https://doi.org/10.1063/5.0037995} {\bibfield  {journal} {\bibinfo
  {journal} {J. Chem. Phys.}\ }\textbf {\bibinfo {volume} {154}},\ \bibinfo
  {pages} {094120} (\bibinfo {year} {2021})}\BibitemShut {NoStop}%
\bibitem [{\citenamefont {Fischer}\ and\ \citenamefont
  {Saalfrank}(2021)}]{21FiSa}%
  \BibitemOpen
  \bibfield  {author} {\bibinfo {author} {\bibfnamefont {E.~W.}\ \bibnamefont
  {Fischer}}\ and\ \bibinfo {author} {\bibfnamefont {P.}~\bibnamefont
  {Saalfrank}},\ }\bibfield  {title} {\enquote {\bibinfo {title} {Ground state
  properties and infrared spectra of anharmonic vibrational polaritons of small
  molecules in cavities},}\ }\href {https://doi.org/10.1063/5.0040853}
  {\bibfield  {journal} {\bibinfo  {journal} {J. Chem. Phys.}\ }\textbf
  {\bibinfo {volume} {154}},\ \bibinfo {pages} {104311} (\bibinfo {year}
  {2021})}\BibitemShut {NoStop}%
\bibitem [{\citenamefont {Farag}, \citenamefont {Mandal},\ and\ \citenamefont
  {Huo}(2021)}]{D1CP00943E}%
  \BibitemOpen
  \bibfield  {author} {\bibinfo {author} {\bibfnamefont {M.~H.}\ \bibnamefont
  {Farag}}, \bibinfo {author} {\bibfnamefont {A.}~\bibnamefont {Mandal}},\ and\
  \bibinfo {author} {\bibfnamefont {P.}~\bibnamefont {Huo}},\ }\bibfield
  {title} {\enquote {\bibinfo {title} {Polariton induced conical intersection
  and berry phase},}\ }\href {https://doi.org/10.1039/D1CP00943E} {\bibfield
  {journal} {\bibinfo  {journal} {Phys. Chem. Chem. Phys.}\ }\textbf {\bibinfo
  {volume} {23}},\ \bibinfo {pages} {16868--16879} (\bibinfo {year}
  {2021})}\BibitemShut {NoStop}%
\bibitem [{\citenamefont {F{\'{a}}bri}, \citenamefont {Hal{\'{a}}sz},\ and\
  \citenamefont {Vib{\'{o}}k}(2022)}]{Fbri2022}%
  \BibitemOpen
  \bibfield  {author} {\bibinfo {author} {\bibfnamefont {C.}~\bibnamefont
  {F{\'{a}}bri}}, \bibinfo {author} {\bibfnamefont {G.~J.}\ \bibnamefont
  {Hal{\'{a}}sz}},\ and\ \bibinfo {author} {\bibfnamefont
  {{\'{A}}.}~\bibnamefont {Vib{\'{o}}k}},\ }\bibfield  {title} {\enquote
  {\bibinfo {title} {Probing light-induced conical intersections by monitoring
  multidimensional polaritonic surfaces},}\ }\href
  {https://doi.org/10.1021/acs.jpclett.1c03465} {\bibfield  {journal} {\bibinfo
   {journal} {J. Phys. Chem. Lett.}\ }\textbf {\bibinfo {volume} {13}},\
  \bibinfo {pages} {1172--1179} (\bibinfo {year} {2022})}\BibitemShut {NoStop}%
\bibitem [{\citenamefont {Sch\"afer}(2022)}]{Schfer2022}%
  \BibitemOpen
  \bibfield  {author} {\bibinfo {author} {\bibfnamefont {C.}~\bibnamefont
  {Sch\"afer}},\ }\bibfield  {title} {\enquote {\bibinfo {title} {Polaritonic
  chemistry from first principles via embedding radiation reaction},}\ }\href
  {https://doi.org/10.1021/acs.jpclett.2c01169} {\bibfield  {journal} {\bibinfo
   {journal} {J. Phys. Chem. Lett.}\ }\textbf {\bibinfo {volume} {13}},\
  \bibinfo {pages} {6905--6911} (\bibinfo {year} {2022})}\BibitemShut {NoStop}%
\bibitem [{\citenamefont {Fischer}\ and\ \citenamefont
  {Saalfrank}(2022)}]{Fischer2022}%
  \BibitemOpen
  \bibfield  {author} {\bibinfo {author} {\bibfnamefont {E.~W.}\ \bibnamefont
  {Fischer}}\ and\ \bibinfo {author} {\bibfnamefont {P.}~\bibnamefont
  {Saalfrank}},\ }\bibfield  {title} {\enquote {\bibinfo {title}
  {Cavity-induced non-adiabatic dynamics and spectroscopy of molecular
  rovibrational polaritons studied by multi-mode quantum models},}\ }\href
  {https://doi.org/10.1063/5.0098006} {\bibfield  {journal} {\bibinfo
  {journal} {J. Chem. Phys.}\ }\textbf {\bibinfo {volume} {157}},\ \bibinfo
  {pages} {034305} (\bibinfo {year} {2022})}\BibitemShut {NoStop}%
\bibitem [{\citenamefont {Li}\ \emph {et~al.}(2022)\citenamefont {Li},
  \citenamefont {Cui}, \citenamefont {Subotnik},\ and\ \citenamefont
  {Nitzan}}]{Li2022}%
  \BibitemOpen
  \bibfield  {author} {\bibinfo {author} {\bibfnamefont {T.~E.}\ \bibnamefont
  {Li}}, \bibinfo {author} {\bibfnamefont {B.}~\bibnamefont {Cui}}, \bibinfo
  {author} {\bibfnamefont {J.~E.}\ \bibnamefont {Subotnik}},\ and\ \bibinfo
  {author} {\bibfnamefont {A.}~\bibnamefont {Nitzan}},\ }\bibfield  {title}
  {\enquote {\bibinfo {title} {Molecular polaritonics: Chemical dynamics under
  strong light{\textendash}matter coupling},}\ }\href
  {https://doi.org/10.1146/annurev-physchem-090519-042621} {\bibfield
  {journal} {\bibinfo  {journal} {Annu. Rev. Phys. Chem.}\ }\textbf {\bibinfo
  {volume} {73}},\ \bibinfo {pages} {43--71} (\bibinfo {year}
  {2022})}\BibitemShut {NoStop}%
\bibitem [{\citenamefont {Cui}\ and\ \citenamefont {Nitzan}(2022)}]{Cui2022}%
  \BibitemOpen
  \bibfield  {author} {\bibinfo {author} {\bibfnamefont {B.}~\bibnamefont
  {Cui}}\ and\ \bibinfo {author} {\bibfnamefont {A.}~\bibnamefont {Nitzan}},\
  }\bibfield  {title} {\enquote {\bibinfo {title} {Collective response in
  light-matter interactions: The interplay between strong coupling and local
  dynamics},}\ }\href {https://doi.org/10.1063/5.0101528} {\bibfield  {journal}
  {\bibinfo  {journal} {J. Chem. Phys.}\ }\textbf {\bibinfo {volume} {157}},\
  \bibinfo {pages} {114108} (\bibinfo {year} {2022})}\BibitemShut {NoStop}%
\bibitem [{\citenamefont {Wellnitz}, \citenamefont {Pupillo},\ and\
  \citenamefont {Schachenmayer}(2022)}]{Wellnitz2022}%
  \BibitemOpen
  \bibfield  {author} {\bibinfo {author} {\bibfnamefont {D.}~\bibnamefont
  {Wellnitz}}, \bibinfo {author} {\bibfnamefont {G.}~\bibnamefont {Pupillo}},\
  and\ \bibinfo {author} {\bibfnamefont {J.}~\bibnamefont {Schachenmayer}},\
  }\bibfield  {title} {\enquote {\bibinfo {title} {Disorder enhanced
  vibrational entanglement and dynamics in polaritonic chemistry},}\ }\href
  {https://doi.org/10.1038/s42005-022-00892-5} {\bibfield  {journal} {\bibinfo
  {journal} {Commun. Phys.}\ }\textbf {\bibinfo {volume} {5}},\ \bibinfo
  {pages} {120} (\bibinfo {year} {2022})}\BibitemShut {NoStop}%
\bibitem [{\citenamefont {Malave}\ \emph {et~al.}(2022)\citenamefont {Malave},
  \citenamefont {Aklilu}, \citenamefont {Beutel}, \citenamefont {Huang},\ and\
  \citenamefont {Varga}}]{Malave2022}%
  \BibitemOpen
  \bibfield  {author} {\bibinfo {author} {\bibfnamefont {J.}~\bibnamefont
  {Malave}}, \bibinfo {author} {\bibfnamefont {Y.~S.}\ \bibnamefont {Aklilu}},
  \bibinfo {author} {\bibfnamefont {M.}~\bibnamefont {Beutel}}, \bibinfo
  {author} {\bibfnamefont {C.}~\bibnamefont {Huang}},\ and\ \bibinfo {author}
  {\bibfnamefont {K.}~\bibnamefont {Varga}},\ }\bibfield  {title} {\enquote
  {\bibinfo {title} {Harmonically confined {$N$-electron} systems coupled to
  light in a cavity},}\ }\href {https://doi.org/10.1103/PhysRevB.105.115127}
  {\bibfield  {journal} {\bibinfo  {journal} {Phys. Rev. B}\ }\textbf {\bibinfo
  {volume} {105}},\ \bibinfo {pages} {115127} (\bibinfo {year}
  {2022})}\BibitemShut {NoStop}%
\bibitem [{\citenamefont {Reitz}, \citenamefont {Sommer},\ and\ \citenamefont
  {Genes}(2022)}]{22ReSoGe}%
  \BibitemOpen
  \bibfield  {author} {\bibinfo {author} {\bibfnamefont {M.}~\bibnamefont
  {Reitz}}, \bibinfo {author} {\bibfnamefont {C.}~\bibnamefont {Sommer}},\ and\
  \bibinfo {author} {\bibfnamefont {C.}~\bibnamefont {Genes}},\ }\bibfield
  {title} {\enquote {\bibinfo {title} {Cooperative quantum phenomena in
  light-matter platforms},}\ }\href
  {https://doi.org/10.1103/PRXQuantum.3.010201} {\bibfield  {journal} {\bibinfo
   {journal} {PRX Quantum}\ }\textbf {\bibinfo {volume} {3}},\ \bibinfo {pages}
  {010201} (\bibinfo {year} {2022})}\BibitemShut {NoStop}%
\bibitem [{\citenamefont {Fregoni}, \citenamefont {Garcia-Vidal},\ and\
  \citenamefont {Feist}(2022)}]{Fregoni2022}%
  \BibitemOpen
  \bibfield  {author} {\bibinfo {author} {\bibfnamefont {J.}~\bibnamefont
  {Fregoni}}, \bibinfo {author} {\bibfnamefont {F.~J.}\ \bibnamefont
  {Garcia-Vidal}},\ and\ \bibinfo {author} {\bibfnamefont {J.}~\bibnamefont
  {Feist}},\ }\bibfield  {title} {\enquote {\bibinfo {title} {Theoretical
  challenges in polaritonic chemistry},}\ }\href
  {https://doi.org/10.1021/acsphotonics.1c01749} {\bibfield  {journal}
  {\bibinfo  {journal} {ACS Photonics}\ }\textbf {\bibinfo {volume} {9}},\
  \bibinfo {pages} {1096--1107} (\bibinfo {year} {2022})}\BibitemShut {NoStop}%
\bibitem [{\citenamefont {Bitton}\ and\ \citenamefont {Haran}(2022)}]{22BiHa}%
  \BibitemOpen
  \bibfield  {author} {\bibinfo {author} {\bibfnamefont {O.}~\bibnamefont
  {Bitton}}\ and\ \bibinfo {author} {\bibfnamefont {G.}~\bibnamefont {Haran}},\
  }\bibfield  {title} {\enquote {\bibinfo {title} {Plasmonic cavities and
  individual quantum emitters in the strong coupling limit},}\ }\href
  {https://doi.org/10.1021/acs.accounts.2c00028} {\bibfield  {journal}
  {\bibinfo  {journal} {Acc. Chem. Res.}\ }\textbf {\bibinfo {volume} {55}},\
  \bibinfo {pages} {1659--1668} (\bibinfo {year} {2022})}\BibitemShut {NoStop}%
\bibitem [{\citenamefont {Fan}\ \emph {et~al.}(2023)\citenamefont {Fan},
  \citenamefont {Shu}, \citenamefont {Dong}, \citenamefont {He}, \citenamefont
  {Henriksen},\ and\ \citenamefont {Nori}}]{Fan2023}%
  \BibitemOpen
  \bibfield  {author} {\bibinfo {author} {\bibfnamefont {L.-B.}\ \bibnamefont
  {Fan}}, \bibinfo {author} {\bibfnamefont {C.-C.}\ \bibnamefont {Shu}},
  \bibinfo {author} {\bibfnamefont {D.}~\bibnamefont {Dong}}, \bibinfo {author}
  {\bibfnamefont {J.}~\bibnamefont {He}}, \bibinfo {author} {\bibfnamefont
  {N.~E.}\ \bibnamefont {Henriksen}},\ and\ \bibinfo {author} {\bibfnamefont
  {F.}~\bibnamefont {Nori}},\ }\bibfield  {title} {\enquote {\bibinfo {title}
  {Quantum coherent control of a single molecular-polariton rotation},}\ }\href
  {https://doi.org/10.1103/PhysRevLett.130.043604} {\bibfield  {journal}
  {\bibinfo  {journal} {Phys. Rev. Lett.}\ }\textbf {\bibinfo {volume} {130}},\
  \bibinfo {pages} {043604} (\bibinfo {year} {2023})}\BibitemShut {NoStop}%
\bibitem [{\citenamefont {Schnappinger}\ and\ \citenamefont
  {Kowalewski}(2023)}]{23ScKo}%
  \BibitemOpen
  \bibfield  {author} {\bibinfo {author} {\bibfnamefont {T.}~\bibnamefont
  {Schnappinger}}\ and\ \bibinfo {author} {\bibfnamefont {M.}~\bibnamefont
  {Kowalewski}},\ }\bibfield  {title} {\enquote {\bibinfo {title} {Nonadiabatic
  wave packet dynamics with ab initio cavity-born-oppenheimer potential energy
  surfaces},}\ }\href {https://doi.org/10.1021/acs.jctc.2c01154} {\bibfield
  {journal} {\bibinfo  {journal} {J. Chem. Theory Comput.}\ }\textbf {\bibinfo
  {volume} {19}},\ \bibinfo {pages} {460--471} (\bibinfo {year}
  {2023})}\BibitemShut {NoStop}%
\bibitem [{\citenamefont {Koner}\ \emph {et~al.}(2023)\citenamefont {Koner},
  \citenamefont {Du}, \citenamefont {Pannir-Sivajothi}, \citenamefont
  {Goldsmith},\ and\ \citenamefont {Yuen-Zhou}}]{23KoDuPa}%
  \BibitemOpen
  \bibfield  {author} {\bibinfo {author} {\bibfnamefont {A.}~\bibnamefont
  {Koner}}, \bibinfo {author} {\bibfnamefont {M.}~\bibnamefont {Du}}, \bibinfo
  {author} {\bibfnamefont {S.}~\bibnamefont {Pannir-Sivajothi}}, \bibinfo
  {author} {\bibfnamefont {R.~H.}\ \bibnamefont {Goldsmith}},\ and\ \bibinfo
  {author} {\bibfnamefont {J.}~\bibnamefont {Yuen-Zhou}},\ }\bibfield  {title}
  {\enquote {\bibinfo {title} {A path towards single molecule vibrational
  strong coupling in a {Fabry-P\'erot} microcavity},}\ }\href
  {https://doi.org/10.1039/D3SC01411H} {\bibfield  {journal} {\bibinfo
  {journal} {Chem. Sci.}\ }\textbf {\bibinfo {volume} {14}},\ \bibinfo {pages}
  {7753--7761} (\bibinfo {year} {2023})}\BibitemShut {NoStop}%
\bibitem [{\citenamefont {Sch\"afer}\ \emph {et~al.}(2020)\citenamefont
  {Sch\"afer}, \citenamefont {Ruggenthaler}, \citenamefont {Rokaj},\ and\
  \citenamefont {Rubio}}]{20ScRuRo}%
  \BibitemOpen
  \bibfield  {author} {\bibinfo {author} {\bibfnamefont {C.}~\bibnamefont
  {Sch\"afer}}, \bibinfo {author} {\bibfnamefont {M.}~\bibnamefont
  {Ruggenthaler}}, \bibinfo {author} {\bibfnamefont {V.}~\bibnamefont
  {Rokaj}},\ and\ \bibinfo {author} {\bibfnamefont {A.}~\bibnamefont {Rubio}},\
  }\bibfield  {title} {\enquote {\bibinfo {title} {Relevance of the quadratic
  diamagnetic and self-polarization terms in cavity quantum electrodynamics},}\
  }\href {https://doi.org/10.1021/acsphotonics.9b01649} {\bibfield  {journal}
  {\bibinfo  {journal} {ACS Photonics}\ }\textbf {\bibinfo {volume} {7}},\
  \bibinfo {pages} {975--990} (\bibinfo {year} {2020})}\BibitemShut {NoStop}%
\bibitem [{\citenamefont {Mandal}, \citenamefont {Montillo~Vega},\ and\
  \citenamefont {Huo}(2020)}]{20MaMoHu}%
  \BibitemOpen
  \bibfield  {author} {\bibinfo {author} {\bibfnamefont {A.}~\bibnamefont
  {Mandal}}, \bibinfo {author} {\bibfnamefont {S.}~\bibnamefont
  {Montillo~Vega}},\ and\ \bibinfo {author} {\bibfnamefont {P.}~\bibnamefont
  {Huo}},\ }\bibfield  {title} {\enquote {\bibinfo {title} {Polarized {Fock}
  states and the dynamical {Casimir} effect in molecular cavity quantum
  electrodynamics},}\ }\href {https://doi.org/10.1021/acs.jpclett.0c02399}
  {\bibfield  {journal} {\bibinfo  {journal} {J. Phys. Chem. Lett.}\ }\textbf
  {\bibinfo {volume} {11}},\ \bibinfo {pages} {9215--9223} (\bibinfo {year}
  {2020})}\BibitemShut {NoStop}%
\bibitem [{\citenamefont {Taylor}\ \emph {et~al.}(2020)\citenamefont {Taylor},
  \citenamefont {Mandal}, \citenamefont {Zhou},\ and\ \citenamefont
  {Huo}}]{20TaMaZh}%
  \BibitemOpen
  \bibfield  {author} {\bibinfo {author} {\bibfnamefont {M.~A.~D.}\
  \bibnamefont {Taylor}}, \bibinfo {author} {\bibfnamefont {A.}~\bibnamefont
  {Mandal}}, \bibinfo {author} {\bibfnamefont {W.}~\bibnamefont {Zhou}},\ and\
  \bibinfo {author} {\bibfnamefont {P.}~\bibnamefont {Huo}},\ }\bibfield
  {title} {\enquote {\bibinfo {title} {Resolution of gauge ambiguities in
  molecular cavity quantum electrodynamics},}\ }\href
  {https://doi.org/10.1103/PhysRevLett.125.123602} {\bibfield  {journal}
  {\bibinfo  {journal} {Phys. Rev. Lett.}\ }\textbf {\bibinfo {volume} {125}},\
  \bibinfo {pages} {123602} (\bibinfo {year} {2020})}\BibitemShut {NoStop}%
\bibitem [{\citenamefont {Galego}, \citenamefont {Garcia-Vidal},\ and\
  \citenamefont {Feist}(2016)}]{Galego2016}%
  \BibitemOpen
  \bibfield  {author} {\bibinfo {author} {\bibfnamefont {J.}~\bibnamefont
  {Galego}}, \bibinfo {author} {\bibfnamefont {F.~J.}\ \bibnamefont
  {Garcia-Vidal}},\ and\ \bibinfo {author} {\bibfnamefont {J.}~\bibnamefont
  {Feist}},\ }\bibfield  {title} {\enquote {\bibinfo {title} {Suppressing
  photochemical reactions with quantized light fields},}\ }\href
  {https://doi.org/10.1038/ncomms13841} {\bibfield  {journal} {\bibinfo
  {journal} {Nat. Commun.}\ }\textbf {\bibinfo {volume} {7}},\ \bibinfo {pages}
  {13841} (\bibinfo {year} {2016})}\BibitemShut {NoStop}%
\bibitem [{\citenamefont {Szidarovszky}, \citenamefont {Hal{\'{a}}sz},\ and\
  \citenamefont {Vib{\'{o}}k}(2020)}]{Szidarovszky2020}%
  \BibitemOpen
  \bibfield  {author} {\bibinfo {author} {\bibfnamefont {T.}~\bibnamefont
  {Szidarovszky}}, \bibinfo {author} {\bibfnamefont {G.~J.}\ \bibnamefont
  {Hal{\'{a}}sz}},\ and\ \bibinfo {author} {\bibfnamefont
  {{\'{A}}.}~\bibnamefont {Vib{\'{o}}k}},\ }\bibfield  {title} {\enquote
  {\bibinfo {title} {Three-player polaritons: Nonadiabatic fingerprints in an
  entangled atom-molecule-photon system},}\ }\href
  {https://doi.org/10.1088/1367-2630/ab8264} {\bibfield  {journal} {\bibinfo
  {journal} {New J. Phys.}\ }\textbf {\bibinfo {volume} {22}},\ \bibinfo
  {pages} {053001} (\bibinfo {year} {2020})}\BibitemShut {NoStop}%
\bibitem [{\citenamefont {F{\'{a}}bri}\ \emph {et~al.}(2022)\citenamefont
  {F{\'{a}}bri}, \citenamefont {Hal{\'{a}}sz}, \citenamefont {Cederbaum},\ and\
  \citenamefont {Vib{\'{o}}k}}]{Fbri2023}%
  \BibitemOpen
  \bibfield  {author} {\bibinfo {author} {\bibfnamefont {C.}~\bibnamefont
  {F{\'{a}}bri}}, \bibinfo {author} {\bibfnamefont {G.~J.}\ \bibnamefont
  {Hal{\'{a}}sz}}, \bibinfo {author} {\bibfnamefont {L.~S.}\ \bibnamefont
  {Cederbaum}},\ and\ \bibinfo {author} {\bibfnamefont {{\'{A}}.}~\bibnamefont
  {Vib{\'{o}}k}},\ }\bibfield  {title} {\enquote {\bibinfo {title} {Radiative
  emission of polaritons controlled by light-induced geometric phase},}\ }\href
  {https://doi.org/10.1039/d2cc04222c} {\bibfield  {journal} {\bibinfo
  {journal} {Chem. Commun.}\ }\textbf {\bibinfo {volume} {58}},\ \bibinfo
  {pages} {12612--12615} (\bibinfo {year} {2022})}\BibitemShut {NoStop}%
\bibitem [{\citenamefont {Csehi}\ \emph {et~al.}(2022)\citenamefont {Csehi},
  \citenamefont {Vendrell}, \citenamefont {Hal{\'{a}}sz},\ and\ \citenamefont
  {Vib{\'{o}}k}}]{Csehi2022}%
  \BibitemOpen
  \bibfield  {author} {\bibinfo {author} {\bibfnamefont {A.}~\bibnamefont
  {Csehi}}, \bibinfo {author} {\bibfnamefont {O.}~\bibnamefont {Vendrell}},
  \bibinfo {author} {\bibfnamefont {G.~J.}\ \bibnamefont {Hal{\'{a}}sz}},\ and\
  \bibinfo {author} {\bibfnamefont {{\'{A}}.}~\bibnamefont {Vib{\'{o}}k}},\
  }\bibfield  {title} {\enquote {\bibinfo {title} {Competition between
  collective and individual conical intersection dynamics in an optical
  cavity},}\ }\href {https://doi.org/10.1088/1367-2630/ac7df7} {\bibfield
  {journal} {\bibinfo  {journal} {New J. Phys.}\ }\textbf {\bibinfo {volume}
  {24}},\ \bibinfo {pages} {073022} (\bibinfo {year} {2022})}\BibitemShut
  {NoStop}%
\bibitem [{\citenamefont {K\"oppel}, \citenamefont {Domcke},\ and\
  \citenamefont {Cederbaum}(1984)}]{Koppel198459}%
  \BibitemOpen
  \bibfield  {author} {\bibinfo {author} {\bibfnamefont {H.}~\bibnamefont
  {K\"oppel}}, \bibinfo {author} {\bibfnamefont {W.}~\bibnamefont {Domcke}},\
  and\ \bibinfo {author} {\bibfnamefont {L.~S.}\ \bibnamefont {Cederbaum}},\
  }\bibfield  {title} {\enquote {\bibinfo {title} {Multimode molecular dynamics
  beyond the {Born--Oppenheimer} approximation},}\ }\href
  {https://doi.org/10.1002/9780470142813.ch2} {\bibfield  {journal} {\bibinfo
  {journal} {Adv. Chem. Phys.}\ }\textbf {\bibinfo {volume} {57}},\ \bibinfo
  {pages} {59--246} (\bibinfo {year} {1984})}\BibitemShut {NoStop}%
\bibitem [{\citenamefont {Yarkony}(1996)}]{Yarkony1996985}%
  \BibitemOpen
  \bibfield  {author} {\bibinfo {author} {\bibfnamefont {D.}~\bibnamefont
  {Yarkony}},\ }\bibfield  {title} {\enquote {\bibinfo {title} {Diabolical
  conical intersections},}\ }\href {https://doi.org/10.1103/RevModPhys.68.985}
  {\bibfield  {journal} {\bibinfo  {journal} {Rev. Mod. Phys.}\ }\textbf
  {\bibinfo {volume} {68}},\ \bibinfo {pages} {985--1013} (\bibinfo {year}
  {1996})}\BibitemShut {NoStop}%
\bibitem [{\citenamefont {Baer}(2002)}]{Baer_2002}%
  \BibitemOpen
  \bibfield  {author} {\bibinfo {author} {\bibfnamefont {M.}~\bibnamefont
  {Baer}},\ }\bibfield  {title} {\enquote {\bibinfo {title} {Introduction to
  the theory of electronic non-adiabatic coupling terms in molecular
  systems},}\ }\href {https://doi.org/10.1016/s0370-1573(01)00052-7} {\bibfield
   {journal} {\bibinfo  {journal} {Phys. Rep.}\ }\textbf {\bibinfo {volume}
  {358}},\ \bibinfo {pages} {75--142} (\bibinfo {year} {2002})}\BibitemShut
  {NoStop}%
\bibitem [{\citenamefont {Domcke}, \citenamefont {Yarkony},\ and\ \citenamefont
  {K\"oppel}(2004)}]{Domcke2004}%
  \BibitemOpen
  \bibfield  {author} {\bibinfo {author} {\bibfnamefont {W.}~\bibnamefont
  {Domcke}}, \bibinfo {author} {\bibfnamefont {D.~R.}\ \bibnamefont
  {Yarkony}},\ and\ \bibinfo {author} {\bibfnamefont {H.}~\bibnamefont
  {K\"oppel}},\ }\href {https://doi.org/10.1142/5406} {\emph {\bibinfo {title}
  {Conical Intersections: Electronic Structure, Dynamics and Spectroscopy}}}\
  (\bibinfo  {publisher} {World Scientific, Singapore},\ \bibinfo {year}
  {2004})\BibitemShut {NoStop}%
\bibitem [{\citenamefont {Worth}\ and\ \citenamefont
  {Cederbaum}(2004)}]{Worth2004127}%
  \BibitemOpen
  \bibfield  {author} {\bibinfo {author} {\bibfnamefont {G.}~\bibnamefont
  {Worth}}\ and\ \bibinfo {author} {\bibfnamefont {L.}~\bibnamefont
  {Cederbaum}},\ }\bibfield  {title} {\enquote {\bibinfo {title} {Beyond
  born-oppenheimer: Molecular dynamics through a conical intersection},}\
  }\href {https://doi.org/10.1146/annurev.physchem.55.091602.094335} {\bibfield
   {journal} {\bibinfo  {journal} {Annu. Rev. Phys. Chem.}\ }\textbf {\bibinfo
  {volume} {55}},\ \bibinfo {pages} {127--158} (\bibinfo {year}
  {2004})}\BibitemShut {NoStop}%
\bibitem [{\citenamefont {Baer}(2006)}]{Baer2006}%
  \BibitemOpen
  \bibfield  {author} {\bibinfo {author} {\bibfnamefont {M.}~\bibnamefont
  {Baer}},\ }\href {https://doi.org/10.1002/0471780081} {\emph {\bibinfo
  {title} {Beyond {Born-Oppenheimer}: Electronic Nonadiabatic Coupling Terms
  and Conical Intersections}}}\ (\bibinfo  {publisher} {John Wiley \& Sons:
  Hoboken, NJ},\ \bibinfo {year} {2006})\BibitemShut {NoStop}%
\bibitem [{\citenamefont {\v{S}indelka}, \citenamefont {Moiseyev},\ and\
  \citenamefont {Cederbaum}(2011)}]{Sindelka2011}%
  \BibitemOpen
  \bibfield  {author} {\bibinfo {author} {\bibfnamefont {M.}~\bibnamefont
  {\v{S}indelka}}, \bibinfo {author} {\bibfnamefont {N.}~\bibnamefont
  {Moiseyev}},\ and\ \bibinfo {author} {\bibfnamefont {L.~S.}\ \bibnamefont
  {Cederbaum}},\ }\bibfield  {title} {\enquote {\bibinfo {title} {Strong impact
  of light-induced conical intersections on the spectrum of diatomic
  molecules},}\ }\href {https://doi.org/10.1088/0953-4075/44/4/045603}
  {\bibfield  {journal} {\bibinfo  {journal} {J. Phys. B: At. Mol. Opt. Phys.}\
  }\textbf {\bibinfo {volume} {44}},\ \bibinfo {pages} {045603} (\bibinfo
  {year} {2011})}\BibitemShut {NoStop}%
\bibitem [{\citenamefont {Hal\'asz}, \citenamefont {Vib\'ok},\ and\
  \citenamefont {Cederbaum}(2015)}]{Halasz2015348}%
  \BibitemOpen
  \bibfield  {author} {\bibinfo {author} {\bibfnamefont {G.}~\bibnamefont
  {Hal\'asz}}, \bibinfo {author} {\bibfnamefont {{\'A}.}~\bibnamefont
  {Vib\'ok}},\ and\ \bibinfo {author} {\bibfnamefont {L.}~\bibnamefont
  {Cederbaum}},\ }\bibfield  {title} {\enquote {\bibinfo {title} {Direct
  signature of light-induced conical intersections in diatomics},}\ }\href
  {https://doi.org/10.1021/jz502468d} {\bibfield  {journal} {\bibinfo
  {journal} {J. Phys. Chem. Lett.}\ }\textbf {\bibinfo {volume} {6}},\ \bibinfo
  {pages} {348--354} (\bibinfo {year} {2015})}\BibitemShut {NoStop}%
\bibitem [{\citenamefont {Badank{\'o}}\ \emph {et~al.}(2022)\citenamefont
  {Badank{\'o}}, \citenamefont {Umarov}, \citenamefont {F{\'a}bri},
  \citenamefont {Hal{\'a}sz},\ and\ \citenamefont {Vib{\'o}k}}]{Badanko2022}%
  \BibitemOpen
  \bibfield  {author} {\bibinfo {author} {\bibfnamefont {P.}~\bibnamefont
  {Badank{\'o}}}, \bibinfo {author} {\bibfnamefont {O.}~\bibnamefont {Umarov}},
  \bibinfo {author} {\bibfnamefont {C.}~\bibnamefont {F{\'a}bri}}, \bibinfo
  {author} {\bibfnamefont {G.}~\bibnamefont {Hal{\'a}sz}},\ and\ \bibinfo
  {author} {\bibfnamefont {{\'A}.}~\bibnamefont {Vib{\'o}k}},\ }\bibfield
  {title} {\enquote {\bibinfo {title} {Topological aspects of cavity-induced
  degeneracies in polyatomic molecules},}\ }\href
  {https://doi.org/https://doi.org/10.1002/qua.26750} {\bibfield  {journal}
  {\bibinfo  {journal} {Int. J. Quantum Chem.}\ }\textbf {\bibinfo {volume}
  {122}},\ \bibinfo {pages} {e26750} (\bibinfo {year} {2022})}\BibitemShut
  {NoStop}%
\bibitem [{\citenamefont {DePrince~III}(2021)}]{21DePrince}%
  \BibitemOpen
  \bibfield  {author} {\bibinfo {author} {\bibfnamefont {A.~E.}\ \bibnamefont
  {DePrince~III}},\ }\bibfield  {title} {\enquote {\bibinfo {title}
  {Cavity-modulated ionization potentials and electron affinities from quantum
  electrodynamics coupled-cluster theory},}\ }\href
  {https://doi.org/10.1063/5.0038748} {\bibfield  {journal} {\bibinfo
  {journal} {J. Chem. Phys.}\ }\textbf {\bibinfo {volume} {154}},\ \bibinfo
  {pages} {094112} (\bibinfo {year} {2021})}\BibitemShut {NoStop}%
\bibitem [{\citenamefont {Riso}\ \emph {et~al.}(2022)\citenamefont {Riso},
  \citenamefont {Haugland}, \citenamefont {Ronca},\ and\ \citenamefont
  {Koch}}]{22RiHaRo}%
  \BibitemOpen
  \bibfield  {author} {\bibinfo {author} {\bibfnamefont {R.~R.}\ \bibnamefont
  {Riso}}, \bibinfo {author} {\bibfnamefont {T.~S.}\ \bibnamefont {Haugland}},
  \bibinfo {author} {\bibfnamefont {E.}~\bibnamefont {Ronca}},\ and\ \bibinfo
  {author} {\bibfnamefont {H.}~\bibnamefont {Koch}},\ }\bibfield  {title}
  {\enquote {\bibinfo {title} {On the characteristic features of ionization in
  {QED} environments},}\ }\href {https://doi.org/10.1063/5.0091119} {\bibfield
  {journal} {\bibinfo  {journal} {J. Chem. Phys.}\ }\textbf {\bibinfo {volume}
  {156}},\ \bibinfo {pages} {234103} (\bibinfo {year} {2022})}\BibitemShut
  {NoStop}%
\bibitem [{\citenamefont {Cederbaum}\ \emph {et~al.}(1977)\citenamefont
  {Cederbaum}, \citenamefont {Domcke}, \citenamefont {K\"oppel},\ and\
  \citenamefont {{Von Niessen}}}]{Cederbaum1977}%
  \BibitemOpen
  \bibfield  {author} {\bibinfo {author} {\bibfnamefont {L.}~\bibnamefont
  {Cederbaum}}, \bibinfo {author} {\bibfnamefont {W.}~\bibnamefont {Domcke}},
  \bibinfo {author} {\bibfnamefont {H.}~\bibnamefont {K\"oppel}},\ and\
  \bibinfo {author} {\bibfnamefont {W.}~\bibnamefont {{Von Niessen}}},\
  }\bibfield  {title} {\enquote {\bibinfo {title} {Strong vibronic coupling
  effects in ionization spectra: The ``mystery band'' of butatriene},}\ }\href
  {https://doi.org/https://doi.org/10.1016/0301-0104(77)87041-9} {\bibfield
  {journal} {\bibinfo  {journal} {Chem. Phys.}\ }\textbf {\bibinfo {volume}
  {26}},\ \bibinfo {pages} {169--177} (\bibinfo {year} {1977})}\BibitemShut
  {NoStop}%
\bibitem [{\citenamefont {Cattarius}\ \emph {et~al.}(2001)\citenamefont
  {Cattarius}, \citenamefont {Worth}, \citenamefont {Meyer},\ and\
  \citenamefont {Cederbaum}}]{Cattarius2001}%
  \BibitemOpen
  \bibfield  {author} {\bibinfo {author} {\bibfnamefont {C.}~\bibnamefont
  {Cattarius}}, \bibinfo {author} {\bibfnamefont {G.~A.}\ \bibnamefont
  {Worth}}, \bibinfo {author} {\bibfnamefont {H.-D.}\ \bibnamefont {Meyer}},\
  and\ \bibinfo {author} {\bibfnamefont {L.~S.}\ \bibnamefont {Cederbaum}},\
  }\bibfield  {title} {\enquote {\bibinfo {title} {All mode dynamics at the
  conical intersection of an octa-atomic molecule: Multi-configuration
  time-dependent {Hartree} ({MCTDH}) investigation on the butatriene cation},}\
  }\href {https://doi.org/10.1063/1.1384872} {\bibfield  {journal} {\bibinfo
  {journal} {J. Chem. Phys.}\ }\textbf {\bibinfo {volume} {115}},\ \bibinfo
  {pages} {2088--2100} (\bibinfo {year} {2001})}\BibitemShut {NoStop}%
\bibitem [{\citenamefont {Brogli}\ \emph {et~al.}(1974)\citenamefont {Brogli},
  \citenamefont {Heilbronner}, \citenamefont {Kloster-Jensen}, \citenamefont
  {Schmelzer}, \citenamefont {Manocha}, \citenamefont {Pople},\ and\
  \citenamefont {Radom}}]{Brogli1974}%
  \BibitemOpen
  \bibfield  {author} {\bibinfo {author} {\bibfnamefont {F.}~\bibnamefont
  {Brogli}}, \bibinfo {author} {\bibfnamefont {E.}~\bibnamefont {Heilbronner}},
  \bibinfo {author} {\bibfnamefont {E.}~\bibnamefont {Kloster-Jensen}},
  \bibinfo {author} {\bibfnamefont {A.}~\bibnamefont {Schmelzer}}, \bibinfo
  {author} {\bibfnamefont {A.}~\bibnamefont {Manocha}}, \bibinfo {author}
  {\bibfnamefont {J.}~\bibnamefont {Pople}},\ and\ \bibinfo {author}
  {\bibfnamefont {L.}~\bibnamefont {Radom}},\ }\bibfield  {title} {\enquote
  {\bibinfo {title} {The photoelectron spectrum of butatriene},}\ }\href
  {https://doi.org/https://doi.org/10.1016/0301-0104(74)80051-0} {\bibfield
  {journal} {\bibinfo  {journal} {Chem. Phys.}\ }\textbf {\bibinfo {volume}
  {4}},\ \bibinfo {pages} {107--119} (\bibinfo {year} {1974})}\BibitemShut
  {NoStop}%
\bibitem [{\citenamefont {Cohen-Tannoudji}, \citenamefont {Dupont-Roc},\ and\
  \citenamefont {Grynberg}(2004)}]{04CoDuGr}%
  \BibitemOpen
  \bibfield  {author} {\bibinfo {author} {\bibfnamefont {C.}~\bibnamefont
  {Cohen-Tannoudji}}, \bibinfo {author} {\bibfnamefont {J.}~\bibnamefont
  {Dupont-Roc}},\ and\ \bibinfo {author} {\bibfnamefont {G.}~\bibnamefont
  {Grynberg}},\ }\href@noop {} {\emph {\bibinfo {title} {Atom-Photon
  Interactions: Basic Processes and Applications}}}\ (\bibinfo  {publisher}
  {Wiley-VCH Verlag GmbH and Co. KGaA, Weinheim},\ \bibinfo {year}
  {2004})\BibitemShut {NoStop}%
\bibitem [{\citenamefont {Galego}\ \emph {et~al.}(2019)\citenamefont {Galego},
  \citenamefont {Climent}, \citenamefont {Garcia-Vidal},\ and\ \citenamefont
  {Feist}}]{19GaClGa}%
  \BibitemOpen
  \bibfield  {author} {\bibinfo {author} {\bibfnamefont {J.}~\bibnamefont
  {Galego}}, \bibinfo {author} {\bibfnamefont {C.}~\bibnamefont {Climent}},
  \bibinfo {author} {\bibfnamefont {F.~J.}\ \bibnamefont {Garcia-Vidal}},\ and\
  \bibinfo {author} {\bibfnamefont {J.}~\bibnamefont {Feist}},\ }\bibfield
  {title} {\enquote {\bibinfo {title} {Cavity {Casimir-Polder} forces and their
  effects in ground-state chemical reactivity},}\ }\href
  {https://doi.org/10.1103/PhysRevX.9.021057} {\bibfield  {journal} {\bibinfo
  {journal} {Phys. Rev. X}\ }\textbf {\bibinfo {volume} {9}},\ \bibinfo {pages}
  {021057} (\bibinfo {year} {2019})}\BibitemShut {NoStop}%
\bibitem [{\citenamefont {Feist}, \citenamefont {Fern\'andez-Dom\'inguez},\
  and\ \citenamefont {Garc\'ia-Vidal}(2021)}]{21FeFeGa}%
  \BibitemOpen
  \bibfield  {author} {\bibinfo {author} {\bibfnamefont {J.}~\bibnamefont
  {Feist}}, \bibinfo {author} {\bibfnamefont {A.~I.}\ \bibnamefont
  {Fern\'andez-Dom\'inguez}},\ and\ \bibinfo {author} {\bibfnamefont {F.~J.}\
  \bibnamefont {Garc\'ia-Vidal}},\ }\bibfield  {title} {\enquote {\bibinfo
  {title} {Macroscopic {QED} for quantum nanophotonics: Emitter-centered modes
  as a minimal basis for multiemitter problems},}\ }\href
  {https://doi.org/doi:10.1515/nanoph-2020-0451} {\bibfield  {journal}
  {\bibinfo  {journal} {Nanophotonics}\ }\textbf {\bibinfo {volume} {10}},\
  \bibinfo {pages} {477--489} (\bibinfo {year} {2021})}\BibitemShut {NoStop}%
\bibitem [{\citenamefont {Rokaj}\ \emph {et~al.}(2018)\citenamefont {Rokaj},
  \citenamefont {Welakuh}, \citenamefont {Ruggenthaler},\ and\ \citenamefont
  {Rubio}}]{18RoWeRu}%
  \BibitemOpen
  \bibfield  {author} {\bibinfo {author} {\bibfnamefont {V.}~\bibnamefont
  {Rokaj}}, \bibinfo {author} {\bibfnamefont {D.~M.}\ \bibnamefont {Welakuh}},
  \bibinfo {author} {\bibfnamefont {M.}~\bibnamefont {Ruggenthaler}},\ and\
  \bibinfo {author} {\bibfnamefont {A.}~\bibnamefont {Rubio}},\ }\bibfield
  {title} {\enquote {\bibinfo {title} {Light-matter interaction in the
  long-wavelength limit: No ground-state without dipole self-energy},}\ }\href
  {https://doi.org/10.1088/1361-6455/aa9c99} {\bibfield  {journal} {\bibinfo
  {journal} {J. Phys. B: At. Mol. Opt. Phys.}\ }\textbf {\bibinfo {volume}
  {51}},\ \bibinfo {pages} {034005} (\bibinfo {year} {2018})}\BibitemShut
  {NoStop}%
\bibitem [{\citenamefont {Weichman}\ \emph {et~al.}(2017)\citenamefont
  {Weichman}, \citenamefont {Cheng}, \citenamefont {Kim}, \citenamefont
  {Stanton},\ and\ \citenamefont {Neumark}}]{17WeChKi}%
  \BibitemOpen
  \bibfield  {author} {\bibinfo {author} {\bibfnamefont {M.~L.}\ \bibnamefont
  {Weichman}}, \bibinfo {author} {\bibfnamefont {L.}~\bibnamefont {Cheng}},
  \bibinfo {author} {\bibfnamefont {J.~B.}\ \bibnamefont {Kim}}, \bibinfo
  {author} {\bibfnamefont {J.~F.}\ \bibnamefont {Stanton}},\ and\ \bibinfo
  {author} {\bibfnamefont {D.~M.}\ \bibnamefont {Neumark}},\ }\bibfield
  {title} {\enquote {\bibinfo {title} {Low-lying vibronic level structure of
  the ground state of the methoxy radical: Slow electron velocity-map imaging
  ({SEVI}) spectra and {K\"oppel-Domcke-Cederbaum (KDC)} vibronic {Hamiltonian}
  calculations},}\ }\href {https://doi.org/10.1063/1.4984963} {\bibfield
  {journal} {\bibinfo  {journal} {J. Chem. Phys.}\ }\textbf {\bibinfo {volume}
  {146}},\ \bibinfo {pages} {224309} (\bibinfo {year} {2017})}\BibitemShut
  {NoStop}%
\end{thebibliography}%

\clearpage

\begin{center}
\Large{Supporting Information}
\end{center}

\section{Description of the molecule and group-theoretical considerations}
\label{sec:gt}

\begin{figure}[!hbt]
\includegraphics[scale=0.4]{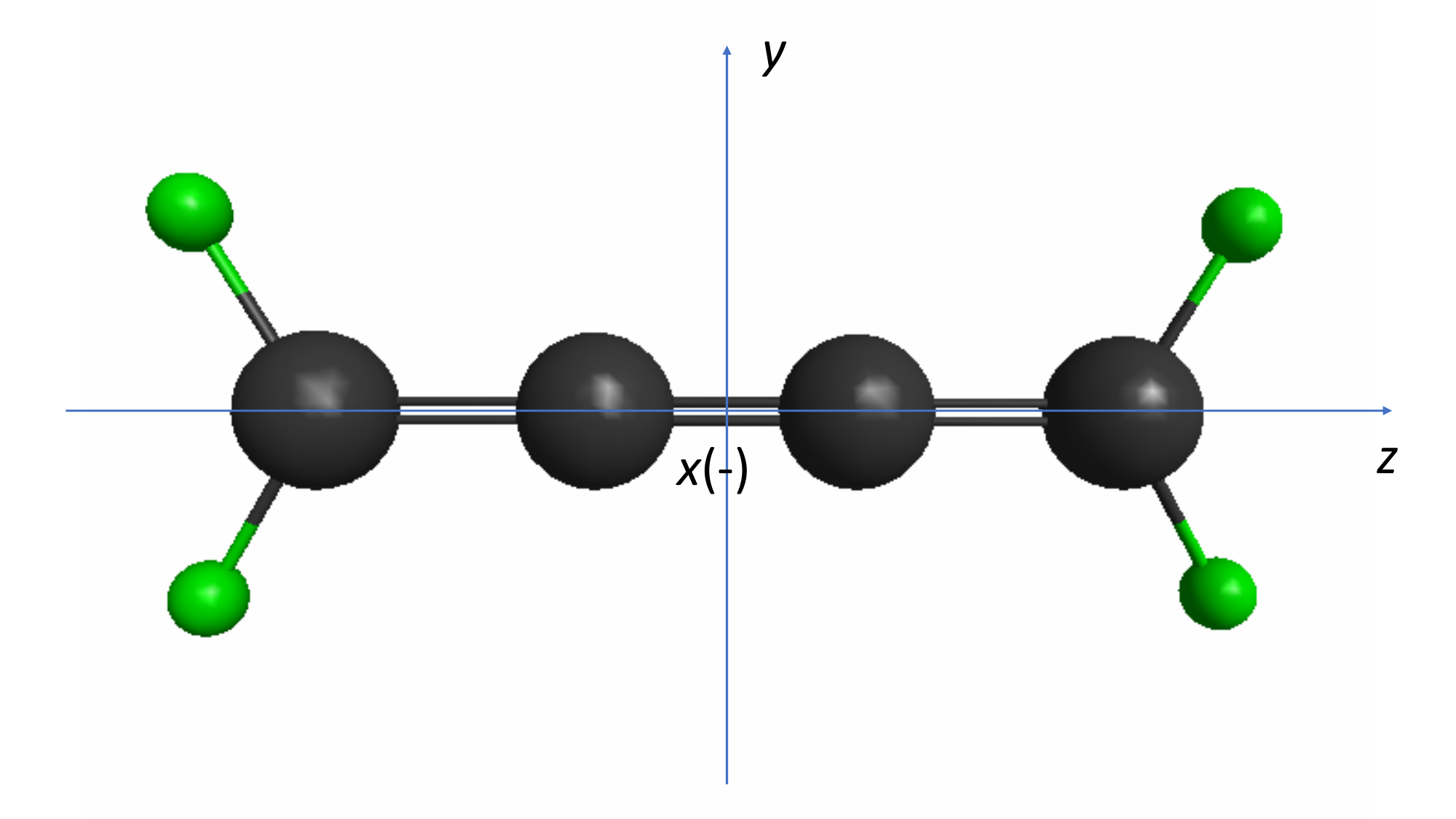}
\caption{\label{fig:c4h4Struc} Equilibrium structure of the neutral butatriene (C$_4$H$_4$) molecule and definition of the body-fixed Cartesian axes.
						The minus sign indicates that the $x$ axis is directed inward.}
\end{figure}

The planar equilibrium structure of butatriene (Franck--Condon (FC) point), shown in Fig. \ref{fig:c4h4Struc}, is of $D_{2\textrm{h}}$ symmetry.
The character table of the $D_{2\textrm{h}}$ point group is
provided in Table \ref{tbl:chtab}. The two lowest electronic states X and A of the cation have $\Gamma_\textrm{X} = B_{2\textrm{g}}$ and
$\Gamma_\textrm{A} = B_{2\textrm{u}}$ symmetries at the FC point, respectively. Following Ref. 70 of the manuscript,
the 18 vibrational modes of butatriene can be classified by $D_{2\textrm{h}}$ irreducible representations (irreps) as
\begin{equation}
	\Gamma_\textrm{vib} = 4 A_\textrm{g} \oplus A_\textrm{u} \oplus 2 B_{2\textrm{g}} \oplus 3 B_{3\textrm{g}} \oplus 3 B_{1\textrm{u}} \oplus 3 B_{2\textrm{u}} \oplus 2 B_{3\textrm{u}}.
   \label{eq:gammaVib}
\end{equation}

\begin{table}
  \caption{Character table of the $D_{2\textrm{h}}$ point group.}
  \label{tbl:chtab}
  \begin{tabular}{rrrrrrrrr}
  \hline
 	& $E$ & $C_2(z)$ & $C_2(y)$ & $C_2(x)$ & $i$ & $\sigma(xy)$ & $\sigma(xz)$ & $\sigma(yz)$ \\ 
  \hline
	$A_\textrm{g}$    & 1 & 1 & 1 & 1 & 1 & 1 & 1 & 1 \\
	$B_{1\textrm{g}}$ & 1 & 1 & $-1$ & $-1$ & 1 & 1 & $-1$ & $-1$ \\
	$B_{2\textrm{g}}$ & 1 & $-1$ & 1 & $-1$ & 1 & $-1$ & 1 & $-1$ \\
	$B_{3\textrm{g}}$ & 1 & $-1$ & $-1$ & 1 & 1 & $-1$ & $-1$ & 1 \\
	$A_\textrm{u}$    & 1 & 1 & 1 & 1 & $-1$ & $-1$ & $-1$ & $-1$ \\
	$B_{1\textrm{u}}$ & 1 & 1 & $-1$ & $-1$ & $-1$ & $-1$ & 1 & 1 \\
	$B_{2\textrm{u}}$ & 1 & $-1$ & 1 & $-1$ & $-1$ & 1 & $-1$ & 1 \\
	$B_{3\textrm{u}}$ & 1 & $-1$ & $-1$ & 1 & $-1$ & 1 & 1 & $-1$ \\
   \hline
\end{tabular}
\end{table}

Using the body-fixed axis definitions in Fig. \ref{fig:c4h4Struc}, we get that the $x$, $y$ and $z$ coordinates
transform according to the irreps
\begin{equation}
	\Gamma_x = B_{3\textrm{u}} \qquad \Gamma_y = B_{2\textrm{u}} \qquad \Gamma_z = B_{1\textrm{u}}.
\end{equation}
Since the direct product representations
\begin{align}
	\Gamma_\textrm{X} \otimes \Gamma_x \otimes \Gamma_\textrm{A}  = B_{2\textrm{g}} \otimes B_{3\textrm{u}} \otimes B_{2\textrm{u}} = B_{3\textrm{g}} \nonumber \\
	\Gamma_\textrm{X} \otimes \Gamma_y \otimes \Gamma_\textrm{A}  = B_{2\textrm{g}} \otimes B_{2\textrm{u}} \otimes B_{2\textrm{u}} = B_{2\textrm{g}} \label{eq:tdmSymm} \\
	\Gamma_\textrm{X} \otimes \Gamma_z \otimes \Gamma_\textrm{A}  = B_{2\textrm{g}} \otimes B_{1\textrm{u}} \otimes B_{2\textrm{u}} = B_{1\textrm{g}} \nonumber
\end{align}
do not contain the totally symmetric irrep $A_\textrm{g}$, all components of the transition dipole moment (TDM) between
the $\textrm{X} ~ ^2B_{2\textrm{g}}$ and $\textrm{A} ~ ^2B_{2\textrm{u}}$ states vanish at the FC point. Using similar arguments one can easily show that
the $\textrm{X} ~ ^2B_{2\textrm{g}}$ and $\textrm{A} ~ ^2B_{2\textrm{u}}$ permanent dipole moments (PDMs) are also zero at the FC point, for example, one gets
\begin{align}
	\Gamma_\textrm{X} \otimes \Gamma_x \otimes \Gamma_\textrm{X}  = B_{2\textrm{g}} \otimes B_{3\textrm{u}} \otimes B_{2\textrm{g}} = B_{3\textrm{u}} \nonumber \\
	\Gamma_\textrm{X} \otimes \Gamma_y \otimes \Gamma_\textrm{X}  = B_{2\textrm{g}} \otimes B_{2\textrm{u}} \otimes B_{2\textrm{g}} = B_{2\textrm{u}} \label{eq:pdmSymm} \\
	\Gamma_\textrm{X} \otimes \Gamma_z \otimes \Gamma_\textrm{X}  = B_{2\textrm{g}} \otimes B_{1\textrm{u}} \otimes B_{2\textrm{g}} = B_{1\textrm{u}} \nonumber
\end{align}
for the $\textrm{X} ~ ^2B_{2\textrm{g}}$ state.

It is evident from Eq. \eqref{eq:tdmSymm} that displacement along a given $B_{3\textrm{g}}$ or $B_{2\textrm{g}}$ vibrational mode
has to be made to produce nonzero TDM along the body-fixed $x$ and $y$ axes, respectively. Since according to Eq. \eqref{eq:gammaVib}
there is no $B_{1\textrm{g}}$ mode, simultaneous displacements along a $B_{3\textrm{g}}$ and a $B_{2\textrm{g}}$ mode are required to
produce TDM along the body-fixed $z$ axis. Similar rules can be derived for PDM components along the same lines, namely,
displacements along $B_{3\textrm{u}}$, $B_{2\textrm{u}}$ and $B_{1\textrm{u}}$ modes generate PDMs along the body-fixed $x$, $y$ and $z$ axes, respectively.

Finally, vibrational modes relevant for the current study are described in Table \ref{tbl:vib}. As discussed in the next section, the modes
$\nu_{14}$ and $\nu_5$ play the role of the tuning ($\nu_\textrm{t}$) and coupling ($\nu_\textrm{c}$) modes, respectively.
In addition to $\nu_\textrm{t}$ and $\nu_\textrm{c}$, we choose modes $\nu_9$ ($\nu_{_\textrm{TDM}}$) and
$\nu_{10}$ ($\nu_{_\textrm{PDM}}$) to generate TDM and PDMs along the $x$ and $y$ axes.
This way, $\nu_{_\textrm{TDM}}$ and $\nu_{_\textrm{PDM}}$ give rise to cavity-molecule couplings upon displacement from the FC point.

\begin{table}
  \caption{Experimental ($\omega_\textrm{exp}$) and calculated ($\omega_\textrm{calc}$, MP2/D95** level of theory) harmonic wavenumbers in units
  		of $\textrm{cm}^{-1}$, symmetry labels ($D_{2\textrm{h}}$ point group) and qualitative descriptions of the
		vibrational modes relevant for the current study. Harmonic wavenumbers used in numerical computations
		are set in bold.}
  \label{tbl:vib}
  \begin{tabular}{ccccc}
 	mode & symmetry & $\omega_\textrm{exp} ~ / ~ \textrm{cm}^{-1}$ & $\omega_\textrm{calc} ~ / ~ \textrm{cm}^{-1}$ & description \\ 
  \hline
	$\nu_{14}$ & $A_\textrm{g}$ & $\mathbf{2079.30}$ & $2157.27$ & C--C stretch (central)  \\
	$\nu_5$ & $A_\textrm{u}$  & $\mathbf{735.58}$ & $777.57$ & torsion \\
	$\nu_9$ & $B_{3\textrm{g}}$  & $662.99$ & $\mathbf{1014.34}$ & CH$_2$ rocking \\
	$\nu_{10}$ & $B_{2\textrm{u}}$  & $\mathbf{1059.81}$ & $1050.12$ & CH$_2$ rocking \\
   \hline
\end{tabular}
\end{table}

\section{Description of the 2-mode vibronic coupling model}

In this section the 2-mode vibronic coupling model is briefly summarized where vibrational modes
$\nu_5$ and $\nu_{14}$ correspond to the coupling ($\nu_\textrm{c}$) and tuning ($\nu_\textrm{t}$) modes, respectively.
The ground electronic state of the neutral molecule is described by the harmonic oscillator Hamiltonian
\begin{equation}
	\hat{H}_0^\textrm{neutral,2D} = \hat{T} + V
	\label{eq:Hneutral}
\end{equation}
with
\begin{equation}
	\hat{T} = -\frac{1}{2} \sum_{i=\textrm{c},\textrm{t}} \frac{\partial^2}{\partial Q_i^2}
\end{equation}
and
\begin{equation}
	V = \frac{1}{2} \sum_{i=\textrm{c},\textrm{t}} \omega_i^2 Q_i^2
\end{equation}
where $Q_\textrm{c}$ and $Q_\textrm{t}$ refer to normal coordinated associated with the coupling and tuning modes, respectively and $\hbar$ is set to one.
The ground (X) and first excited (A) electronic states are taken into account for the cation, which yields the diabatic Hamiltonian
\begin{equation}
	\hat{H}_0^\textrm{ion,2D} = 
	\begin{bmatrix}
            \hat{T} & 0 \\
            0 & \hat{T} 
        \end{bmatrix} + 
        \begin{bmatrix}
            V_\textrm{X} & V_\textrm{XA} \\
            V_\textrm{XA} & V_\textrm{A} 
        \end{bmatrix}
    \label{eq:Hion2D}
\end{equation}
with
\begin{equation}
	V_\textrm{X} = \epsilon_\textrm{X} + \frac{1}{2} \sum_{i=\textrm{c},\textrm{t}} \omega_i^2 Q_i^2 + \kappa_\textrm{X} Q_\textrm{t},
\end{equation}
\begin{equation}
	V_\textrm{A} = \epsilon_\textrm{A} + \frac{1}{2} \sum_{i=\textrm{c},\textrm{t}} \omega_i^2 Q_i^2 + \kappa_\textrm{A} Q_\textrm{t}
\end{equation}
and
\begin{equation}
	V_\textrm{XA} = \lambda Q_\textrm{c}.
   \label{eq:vxa}
\end{equation}
The model parameters ($\epsilon_\textrm{X/A}$, $\omega_i$, $\kappa_\textrm{X/A}$, $\lambda$) have been taken from Ref. 70 of the manuscript.

Next, a numerical method is presented for solving the eigenproblem of $\hat{H}_0^\textrm{ion,2D}$.
The eigenstates of $\hat{H}_0^\textrm{ion,2D}$ are expressed as two-component vectors, that is,
\begin{equation}
	| \vec{\psi} (Q_\textrm{c},Q_\textrm{t}) \rangle =        
	  \begin{bmatrix}
             | \psi_\textrm{X}(Q_\textrm{c},Q_\textrm{t}) \rangle \\
             | \psi_\textrm{A}(Q_\textrm{c},Q_\textrm{t}) \rangle
          \end{bmatrix}
	\label{eq:cation_eigenstate}
\end{equation}
which is equivalent to the expansion 
$| \psi (Q_\textrm{c},Q_\textrm{t}) \rangle = | \psi_\textrm{X}(Q_\textrm{c},Q_\textrm{t}) \rangle | \textrm{X} \rangle + | \psi_\textrm{A}(Q_\textrm{c},Q_\textrm{t}) \rangle | \textrm{A} \rangle$.
Components of $| \vec{\psi}(Q_\textrm{c},Q_\textrm{t}) \rangle$ can be expanded in a product basis of 1D harmonic-oscillator eigenfunctions of the neutral ground state,
\begin{equation}
	| \psi_\alpha(Q_\textrm{c},Q_\textrm{t}) \rangle = \sum_{v_\textrm{c},v_\textrm{t}} c^\alpha_{v_\textrm{c},v_\textrm{t}} | \varphi_{v_\textrm{c}}(Q_\textrm{c}) \rangle | \varphi_{v_\textrm{t}}(Q_\textrm{t}) \rangle
\label{eq:2Dbasis}
\end{equation}
with $\alpha = \textrm{X}, \textrm{A}$. Matrix elements of $\hat{H}_0^\textrm{ion,2D}$ in the basis defined in Eq. \eqref{eq:2Dbasis} read
\begin{gather}
	\langle \varphi_{v'_\textrm{c}} \varphi_{v'_\textrm{t}} | \hat{T} + V_\alpha | \varphi_{v_\textrm{c}} \varphi_{v_\textrm{t}} \rangle =
		\left[ \epsilon_\alpha + \omega_\textrm{c} \left(v_\textrm{c}+\frac{1}{2} \right) +  \omega_\textrm{t} \left(v_\textrm{t}+\frac{1}{2} \right) \right]
		\delta_{v'_\textrm{c},v_\textrm{c}} \delta_{v'_\textrm{t},v_\textrm{t}} \nonumber \\
    + \kappa_\alpha \langle \varphi_{v'_\textrm{t}} | Q_\textrm{t} | \varphi_{v_\textrm{t}} \rangle \delta_{v'_\textrm{c},v_\textrm{c}}
	\label{eq:diag}
\end{gather}
for the diagonal blocks ($\alpha = \textrm{X}, \textrm{A}$) and
\begin{align}
	\langle \varphi_{v'_\textrm{c}} \varphi_{v'_\textrm{t}} | V_\textrm{XA} | \varphi_{v_\textrm{c}} \varphi_{v_\textrm{t}} \rangle =
		\lambda \langle \varphi_{v'_\textrm{c}} | Q_\textrm{c} | \varphi_{v_\textrm{c}} \rangle \delta_{v'_\textrm{t},v_\textrm{t}}
	\label{eq:offdiag}
\end{align}
for the offdiagonal block of $\hat{H}_0^\textrm{ion,2D}$.
In Eq. \eqref{eq:offdiag}, coordinate matrix elements for the coupling mode can be obtained as
\begin{equation}
	\langle \varphi_{v'_\textrm{c}} | Q_\textrm{c} | \varphi_{v_\textrm{c}} \rangle = 
        \frac{1}{\sqrt{2\omega_\textrm{c}}} \left( \sqrt{v_\textrm{c}+1} \, \delta_{v'_\textrm{c},v_{\textrm{c}+1}} + \sqrt{v_\textrm{c}} \, \delta_{v'_\textrm{c},v_{\textrm{c}-1}} \right).
\end{equation}
Coordinate matrix elements for the tuning mode in Eq. \eqref{eq:diag} can be expressed in the same fashion.

Finally, a brief description of the computation of the ionization spectrum is provided. The initial state is assumed to be
the vibrational ground state of the neutral molecule in its electronic ground state,
\begin{equation}
	| \psi_\textrm{i}(Q_\textrm{c},Q_\textrm{t}) \rangle = | \varphi_0(Q_\textrm{c}) \rangle | \varphi_0(Q_\textrm{t}) \rangle
\end{equation}
where the harmonic-oscillator approximation is applied, see Eq. \eqref{eq:Hneutral}.
The final states correspond to eigenstates $| \vec{\psi}_\textrm{f}(Q_\textrm{c},Q_\textrm{t}) \rangle$ of the cation defined in Eq. \eqref{eq:cation_eigenstate}. 
Assuming that ionization probabilities of the neutral ground state are nuclear-position independent and equal for the
cationic states X and A we get 
\begin{align}
	A_\textrm{if} = \sum_{\alpha = \textrm{X}, \textrm{A}} \langle \psi_\textrm{i}(Q_\textrm{c},Q_\textrm{t}) | \psi_{\textrm{f},\alpha}(Q_\textrm{c},Q_\textrm{t}) \rangle
	\label{eq:trampl2D}
\end{align}
for the transition amplitudes.
Transition probabilities $I_\textrm{if}$ are proportional to transition wavenumbers $\omega_\textrm{if}$ and absolute squares of transition amplitudes
$A_\textrm{if}$, that is, $I_\textrm{if} \propto \omega_\textrm{if} |A_\textrm{if}|^2$.

\section{Description of the 4-mode model for the coupled cavity-molecule system}

The Hamiltonian of the cation coupled to a single cavity mode has the form
\begin{equation}
	\hat{H}_\textrm{cm} = \hat{H}_0^\textrm{ion} + \hbar \omega_\textrm{cav} \hat{a}^\dag \hat{a} - g \hat{\vec{\mu}} \vec{e} (\hat{a}^\dag + \hat{a})
\end{equation}
which, taking into account the two lowest electronic states (X and A), can be recast as
\begin{equation}
    \resizebox{0.9\textwidth}{!}{$\hat{H}_\textrm{cm}  = 
         \begin{bmatrix}
            \hat{T} + V_\textrm{X} & V_\textrm{XA} & -g \mu_\textrm{X} & -g \mu_\textrm{XA} & 0 & 0 & \dots \\
            V_\textrm{XA} & \hat{T} + V_\textrm{A} & -g \mu_\textrm{XA} & -g \mu_\textrm{A} & 0 & 0 & \dots \\
            -g \mu_\textrm{X} & -g \mu_\textrm{XA} & \hat{T} + V_\textrm{X} + \hbar\omega_\textrm{cav} & V_\textrm{XA} & -g \sqrt{2} \mu_\textrm{X} & -g \sqrt{2} \mu_\textrm{XA} & \dots \\
            -g \mu_\textrm{XA} & -g \mu_\textrm{A} & V_\textrm{XA} &\hat{T} + V_\textrm{A} + \hbar\omega_\textrm{cav} & -g \sqrt{2} \mu_\textrm{XA} & -g \sqrt{2} \mu_\textrm{A} & \dots \\
            0 & 0 & -g \sqrt{2} \mu_\textrm{X} & -g \sqrt{2} \mu_\textrm{XA} &\hat{T} + V_\textrm{X} + 2\hbar\omega_\textrm{cav} & V_\textrm{XA} & \dots \\
            0 & 0 & -g \sqrt{2} \mu_\textrm{XA} & -g \sqrt{2} \mu_\textrm{A} & V_\textrm{XA} &\hat{T} + V_\textrm{A} + 2\hbar\omega_\textrm{cav} & \dots \\
            \vdots & \vdots & \vdots & \vdots & \vdots & \vdots & \ddots 
        \end{bmatrix}$}
     \label{eq:HcmSI}
\end{equation}
where $\omega_\textrm{cav}$, $g$, $\hat{\vec{\mu}}$ and $\vec{e}$ are the cavity angular frequency, coupling strength parameter, 
electric dipole moment of the molecule and polarization vector of the cavity field, respectively,
while $\hat{a}^\dag$ and $\hat{a}$ denote creation and annihilation operators of the cavity mode.
In addition, PDM and TDM components along $\vec{e}$ are denoted by $\mu_\alpha$ ($\alpha = \textrm{X}, \textrm{A}$) and $\mu_\textrm{XA}$, respectively.

In order to produce nonzero TDM and PDMs, $\hat{H}_0^\textrm{ion,2D}$ of Eq. \eqref{eq:Hion2D} is supplemented with two additional vibrational modes,
which yields
\begin{equation}
	\hat{H}_0^\textrm{ion} = 
	\begin{bmatrix}
            \hat{T} & 0 \\
            0 & \hat{T} 
        \end{bmatrix} + 
        \begin{bmatrix}
            V_\textrm{X} & V_\textrm{XA} \\
            V_\textrm{XA} & V_\textrm{A} 
        \end{bmatrix}
    \label{eq:Hion4D}
\end{equation}
with
\begin{equation}
	\hat{T} = -\frac{1}{2} \sum_{\substack{i=\textrm{c},\textrm{t}, \\ \textrm{TDM},\textrm{PDM}}} \frac{\partial^2}{\partial Q_i^2},
\end{equation}
\begin{equation}
	V_\textrm{X} = \epsilon_\textrm{X} + \frac{1}{2} \sum_{\substack{i=\textrm{c},\textrm{t}, \\ \textrm{TDM},\textrm{PDM}}} \omega_i^2 Q_i^2 + \kappa_\textrm{X} Q_\textrm{t},
\end{equation}
\begin{equation}
	V_\textrm{A} = \epsilon_\textrm{A} + \frac{1}{2} \sum_{\substack{i=\textrm{c},\textrm{t}, \\ \textrm{TDM},\textrm{PDM}}} \omega_i^2 Q_i^2 + \kappa_\textrm{A} Q_\textrm{t}
\end{equation}
and
\begin{equation}
	V_\textrm{XA} = \lambda Q_\textrm{c}.
\end{equation}
The group-theoretical analysis presented in Section \ref{sec:gt} suggests that the PDM and TDM functions can be approximated by the expansions
\begin{align}
	&\mu_\textrm{X} = \beta_\textrm{X} Q_{_\textrm{PDM}} \nonumber \\
	&\mu_\textrm{A} = \beta_\textrm{A} Q_{_\textrm{PDM}} \label{eq:tdmpdm} \\
	&\mu_\textrm{XA} = \gamma Q_{_\textrm{TDM}} \nonumber
\end{align}
where $\beta_\textrm{X}$, $\beta_\textrm{A}$ and $\gamma$ are constants. 

Eigenstates of the coupled cavity-molecule system are constructed as
\begin{equation}
	| \vec{\Phi}_k \rangle =        
	  \begin{bmatrix}
            \sum_\mathbf{v} \sum_{n} b^{\textrm{X},k}_{\mathbf{v},n} | \mathbf{v} \rangle | n \rangle \\
            \sum_\mathbf{v} \sum_{n} b^{\textrm{A},k}_{\mathbf{v},n} | \mathbf{v} \rangle | n \rangle
          \end{bmatrix}
	\label{eq:HcmEigstate}
\end{equation}
where $| n \rangle$ refers to Fock states of the cavity mode
and $| \mathbf{v} \rangle = |v_\textrm{c}, v_\textrm{t}, v_{_\textrm{TDM}}, v_{_\textrm{PDM}} \rangle$ denotes four-dimensional harmonic-oscillator basis states.
The matrix representation of $\hat{H}_\textrm{cm}$ (see Eq. \eqref{eq:HcmSI}) is set up using the basis defined in Eq. \eqref{eq:HcmEigstate}
and the resulting Hamiltonian matrix is diagonalized with a sparse eigensolver.
Regarding the ionization spectrum, the initial state is assumed to be the vibrational ground state of the neutral molecule in
its electronic ground state with zero photons in the cavity, that is,
\begin{equation}
	| \Phi_0 \rangle = | \mathbf{0} \rangle | 0 \rangle.
\end{equation}
With these quantities at hand transition amplitudes from $| \Phi_0 \rangle$ to $| \vec{\Phi}_k \rangle$ (see Eq. \eqref{eq:HcmEigstate}) are expressed as
\begin{align}
	A_{0k} = & \sum_{\alpha = \textrm{X}, \textrm{A}} \langle \Phi_0 | \Phi_{k,\alpha} \rangle.
	\label{eq:tramplCavity}
\end{align}

\section{Technical details of the computations}

\subsection{Ab initio calculations and model parameters}

Following Ref. 70 of the manuscript,
the equilibrium geometry of neutral butatriene was optimized
at the MP2/D95** level of theory with the Gaussian 16 program. Normal coordinates were constructed by diagonalizing
the mass-weighted Cartesian Hessian and used in subsequent nuclear motion computations. Vibrational modes relevant
for the current study are summarized in Table \ref{tbl:vib} where harmonic wavenumbers used in numerical computations
are set in bold. Other model parameters, taken from  Ref. 70 of the manuscript, are as follows:
$\epsilon_\textrm{X} = 9.45 ~ \textrm{eV}$,
$\epsilon_\textrm{A} = 9.85 ~ \textrm{eV}$,
$\kappa_\textrm{X} = -0.212 ~ \textrm{eV}$,
$\kappa_\textrm{A} = 0.255 ~ \textrm{eV}$ and
$\lambda = 0.318 ~ \textrm{eV}$
($\kappa_\textrm{X}$, $\kappa_\textrm{A}$ and $\lambda$ are consistent with dimensionless normal coordinates).

The electronic states X and A of the butatriene cation were obtained at the MRCI/cc-pVTZ level of theory.
First, ROHF orbitals were calculated, then state-averaged CASSCF(5,4) computations were carried out using equal weights
for the two cationic states with 11 closed-shell orbitals and 4 active orbitals. The final step involved the internally-contracted MRCI
method implemented in the Molpro program (version 2020.2). Permanent and transition dipole moments, needed to construct
the Hamiltonian of Eq. \eqref{eq:HcmSI}, were obtained at several different nuclear configurations.

\subsection{Diabatic representation of the permanent and transition dipole moment surfaces}

As the system is described using the diabatic representation, we need to construct the diabatic PDM and TDM surfaces.
This can be achieved by diagonalizing the diabatic potential energy matrix of Eq. \eqref{eq:Hion4D} for each nuclear configuration,
\begin{equation}
	\mathbf{U}^\textrm{T}
        \begin{bmatrix}
            V_\textrm{X} & V_\textrm{XA} \\
            V_\textrm{XA} & V_\textrm{A} 
        \end{bmatrix}
        \mathbf{U} =
        \begin{bmatrix}
            V_\textrm{lower} & 0 \\
            0 & V_\textrm{upper} 
        \end{bmatrix}
    \label{eq:adDiaPot}
\end{equation}
which results in the transformation matrix $\mathbf{U}$ and lower and upper adiabatic PESs ($V_\textrm{lower}$ and $V_\textrm{upper}$).
Since ab initio calculations presented in the previous subsection yield PDM and TDM surfaces in the adiabatic representation, the transformation
\begin{equation}
        \begin{bmatrix}
            \mu_\textrm{X} & \mu_\textrm{XA} \\
            \mu_\textrm{XA} & \mu_\textrm{A} 
        \end{bmatrix} =
         \mathbf{U}
        \begin{bmatrix}
            \mu^\textrm{ad}_\textrm{X} & \mu^\textrm{ad}_\textrm{XA} \\
            \mu^\textrm{ad}_\textrm{XA} & \mu^\textrm{ad}_\textrm{A} 
        \end{bmatrix}
        \mathbf{U}^\textrm{T}
    \label{eq:adDiaDipole}
\end{equation}
is invoked to construct the diabatic PDM ($\mu_\textrm{X}$ and $\mu_\textrm{A}$) and TDM ($\mu_\textrm{XA}$) surfaces.

As already discussed in Section \ref{sec:gt}, the diabatic PDM and TDM surfaces can be approximated by first-order Taylor expansion about the FC point,
\begin{align}
	&\mu_\textrm{X} = \left( \frac{\partial \mu_\textrm{X}}{\partial Q_{_\textrm{PDM}}} \right) \bigg |  _\textrm{FC} Q_{_\textrm{PDM}} = \beta_\textrm{X} Q_{_\textrm{PDM}} \nonumber \\
	&\mu_\textrm{A} = \left( \frac{\partial \mu_\textrm{A}}{\partial Q_{_\textrm{PDM}}} \right) \bigg |  _\textrm{FC} Q_{_\textrm{PDM}} = \beta_\textrm{A} Q_{_\textrm{PDM}} \label{eq:tdmApprox} \\
	&\mu_\textrm{XA} = \left( \frac{\partial \mu_\textrm{XA}}{\partial Q_{_\textrm{TDM}}} \right) \bigg |  _\textrm{FC} Q_{_\textrm{TDM}} = \gamma Q_{_\textrm{TDM}} \nonumber
\end{align}
where PDM and TDM derivatives can be estimated by the method of finite differences. Since $\mat{U}$ equals the two-dimensional identity matrix
in the close vicinity of the FC point, PDM and TDM derivatives in the adiabatic and diabatic representations coincide. 
Therefore, one can directly use dipole derivatives provided by ab initio calculations:
$\beta_\textrm{X} = 1.05 \cdot 10^{-3} ~ \textrm{au}$,
$\beta_\textrm{A} = -1.89 \cdot 10^{-3} ~ \textrm{au}$ and
$\gamma = -4.37 \cdot 10^{-4} ~ \textrm{au}$, given in atomic units.

\section{Additional figures}

\begin{figure}[!hbt]
\includegraphics[scale=0.575]{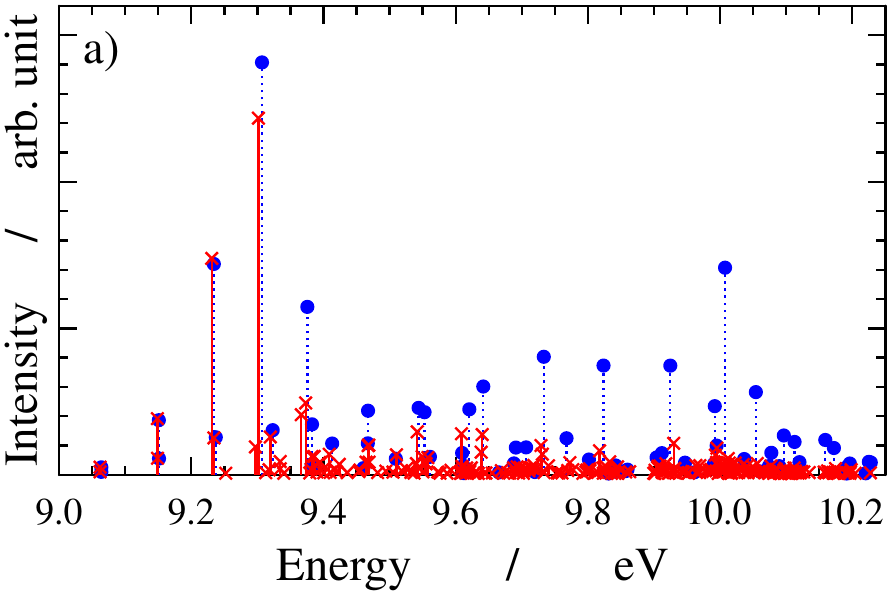}
\includegraphics[scale=0.575]{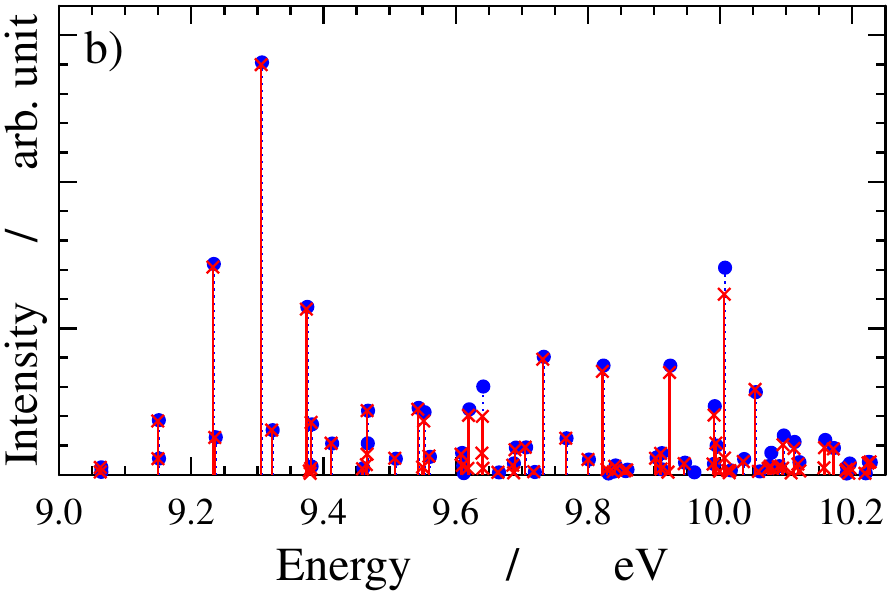}
\includegraphics[scale=0.575]{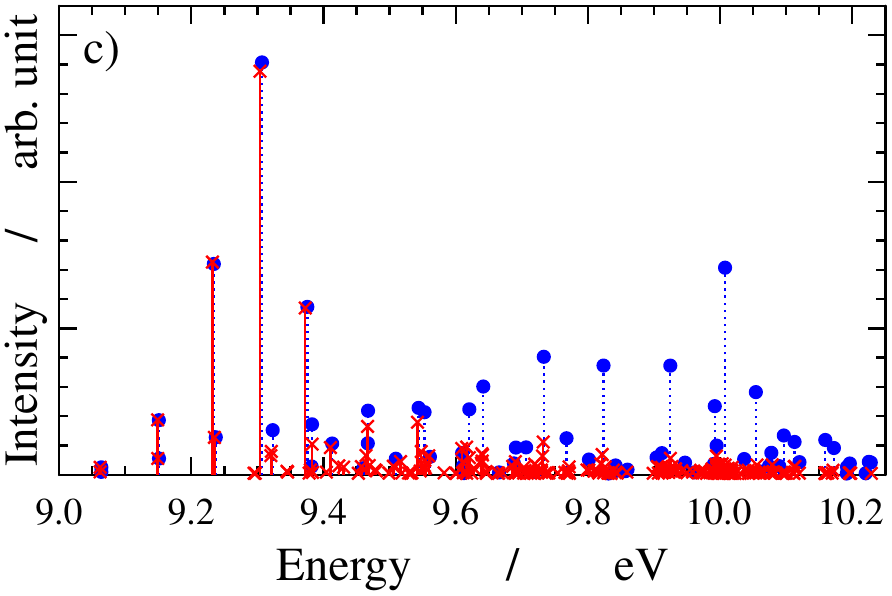}
\caption{\label{fig:spectra1}        
        Cavity-free (blue) and cavity (red) ionization spectra of butatriene with cavity parameters 
        $\omega_\textrm{cav} = 645.02 ~ \textrm{cm}^{-1}$ and $g = 0.15 ~ \textrm{au}$.
        The field polarization vector is chosen in three different ways:
		(a) $\mathbf{e} = (0,1,0)$ (only permanent dipole moments (PDMs)), 
		(b) $\mathbf{e} = (1,0,0)$ (only transition dipole moment (TDM)), and
		(c) $\mathbf{e} = (1,1,0)/\sqrt{2}$ (both TDM and PDMs).}
\end{figure}

\begin{figure}[!hbt]
\includegraphics[scale=0.575]{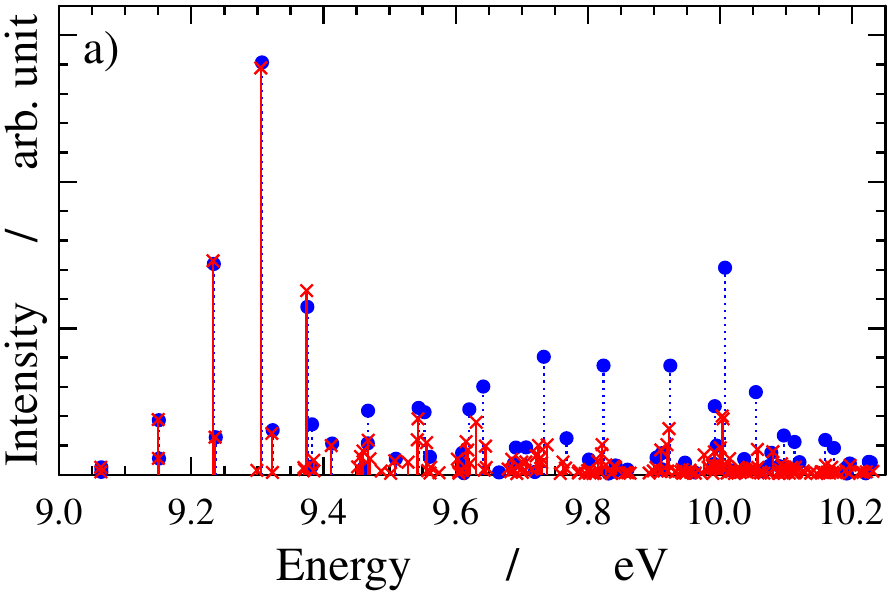}
\includegraphics[scale=0.575]{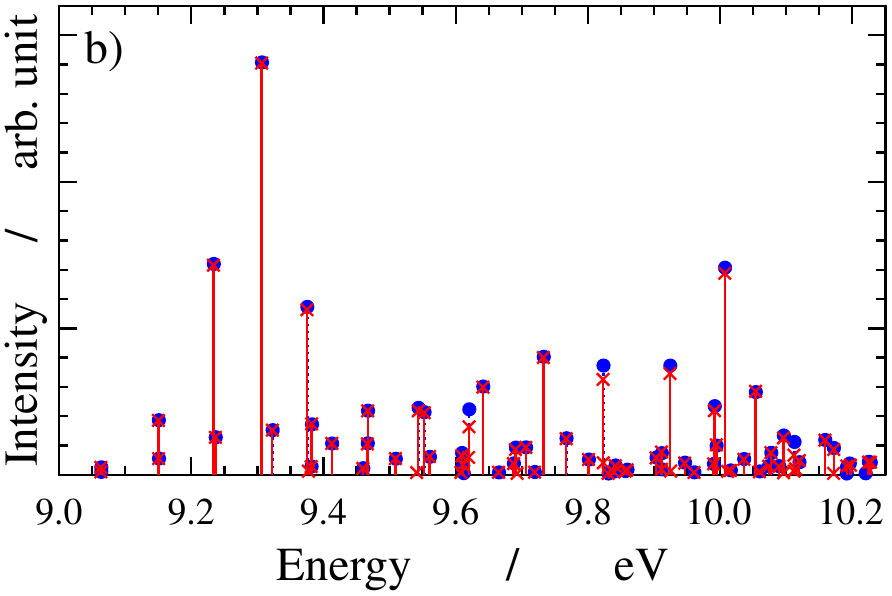}
\includegraphics[scale=0.575]{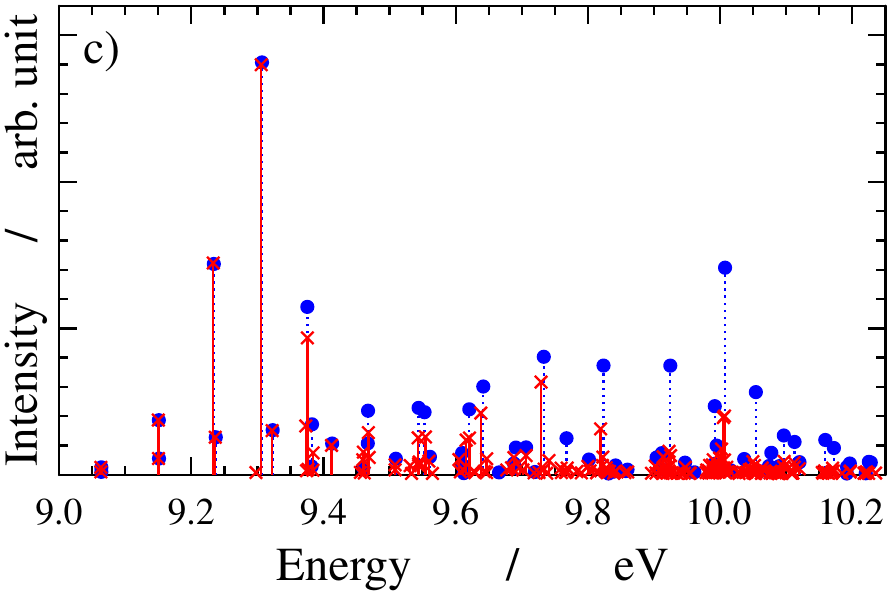}
\caption{\label{fig:spectra2}        
        Cavity-free (blue) and cavity (red) ionization spectra of butatriene with cavity parameters 
        $\omega_\textrm{cav} = 824.80 ~ \textrm{cm}^{-1}$ and $g = 0.1 ~ \textrm{au}$.
        The field polarization vector is chosen in three different ways:
		(a) $\mathbf{e} = (0,1,0)$ (only permanent dipole moments (PDMs)), 
		(b) $\mathbf{e} = (1,0,0)$ (only transition dipole moment (TDM)), and
		(c) $\mathbf{e} = (1,1,0)/\sqrt{2}$ (both TDM and PDMs).}
\end{figure}

Fig. \ref{fig:spectra1} and Fig. \ref{fig:spectra2} show 4-mode cavity-free (blue) and cavity (red) ionization spectra with
$\omega_\textrm{cav} = 645.02 ~ \textrm{cm}^{-1}$ and $g = 0.15 ~ \textrm{au}$,
and $\omega_\textrm{cav} = 824.80 ~ \textrm{cm}^{-1}$ $g = 0.1 ~ \textrm{au}$, respectively.
In both cases, three panels of the figures correspond to the following three different cavity field polarizations
(see Fig. \ref{fig:c4h4Struc} for the definition of body-fixed Cartesian axes):
$\mathbf{e} = (0,1,0)$ (only PDMs), $\mathbf{e} = (1,0,0)$ (only TDM) and $\mathbf{e} = (1,1,0)/\sqrt{2}$ (both TDM and PDMs).

\end{document}